\documentclass[A4paper,11pt]{article}
\usepackage{graphics,graphicx,epsfig,wrapfig}
\usepackage{longtable}  
\usepackage{amsmath,verbatim,amssymb,color,lscape}
\usepackage{mathtools} % 내가 추가함
\usepackage{threeparttable} % 내가 추가함
\usepackage{tabularx} % 내가 추가함
\usepackage{subcaption} % 내가 추가함
\usepackage{url} % 내가 추가함
\usepackage{kotex}
\usepackage{natbib}
\usepackage{lastpage}
\usepackage{multirow} 
\usepackage{booktabs}
\usepackage[table]{xcolor}
\usepackage{booktabs}
\usepackage{authblk}
\usepackage{bm}
\usepackage{indentfirst}
\usepackage[ruled,vlined]{algorithm2e}
\usepackage{algorithmic}
\usepackage[normalem]{ulem}
\usepackage[export]{adjustbox}
\usepackage{verbatim}
\newcommand{\hide}[1]{}

\makeatletter

\usepackage[outerbars,color]{changebar}
\ifx\pdfoutput\undefined
\else\ifnum\pdfoutput>0
  \usepackage{pdfcolmk}
\fi\fi
\cbcolor{black}

\usepackage[margin=1.0in]{geometry}
\usepackage{tikz}
\usetikzlibrary{bayesnet}
\usetikzlibrary{fit,positioning}

\newfont{\rmm}{cmr10 at 11pt}
\rmm

\title{Modeling Transition Dynamics and Network Structure in Cross-National Process Data: A Hierarchical Multi-State Survival Framework}

\author[1]{Doungjun Kim}
\author[1]{Daeun Hwangbo}
\author[3]{Minjeong Jeon}
\author[1,2]{Ick Hoon Jin}
\affil[1]{Department of Statistics and Data Science, Yonsei University. Republic of Korea.}
\affil[2]{Department of Applied Statistics, Yonsei University. Republic of Korea.}
\affil[3]{School of Education and Information Studies, University of California, Los Angeles. USA.}
\date{}

\begin{document}

\begin{titlepage}
\centering

\vspace*{2cm}

{\LARGE \textbf{Modeling Transition Dynamics and Network Structure in Cross-National Process Data: A Hierarchical Multi-State Survival Framework}}

\vspace{2cm}

Doungjun Kim$^1$, Daeun Hwangbo$^1$, Minjeong Jeon$^3$, and Ick Hoon Jin$^{1,2,*}$

\vspace{1.5cm}

$^1$Department of Statistics and Data Science, Yonsei University, Seoul, Republic of Korea

$^2$Department of Applied Statistics, Yonsei University, Seoul, Republic of Korea

$^3$School of Education and Information Studies, University of California, Los Angeles. USA.

\vspace{2cm}

$^*$Corresponding Author:\\
Ick Hoon Jin\\
Department of Applied Statistics,\\ 
Department of Statistics and Data Science,\\
Yonsei University\\
50 Yonsei-ro, Seodaemun-gu, Seoul 03722, Republic of Korea\\
e-mail: ijin@yonsei.ac.kr

\vspace{2cm}

\textbf{Short Running Head:} Cross-National Transition Dynamics in Process Data

\end{titlepage}
\maketitle

\begin{abstract}
Process data from computer-based assessments record the sequence and timing of actions through which respondents solve a task, providing information about both the pace and structure of problem-solving behavior. Modeling such processes across countries is challenging because country-by-response-group cells are often small and unbalanced and the observed transition supports can differ substantially across countries. We propose a hierarchical framework that integrates a Bayesian multi-state survival model with a network-based representation of transition structure. Partial pooling across countries yields country-specific covariate and key-action effects, transition speed, and estimates of between-country heterogeneity. Posterior transition probability networks are embedded in a common latent space using a directed graph auto-encoder adapted to heterogeneous supports, and 1-Wasserstein distances between node-role distributions are evaluated across posterior draws to characterize global network structure while propagating estimation uncertainty. We apply the framework to two problem-solving items from the Programme for the International Assessment of Adult Competencies across 14 countries. The results reveal cross-country heterogeneity in transition speed, systematic response-group differences in global network organization, item-dependent variation in within-group dispersion across countries, and local differences in routing around shared intermediate actions.
\end{abstract}

\newpage
\section{Introduction}\label{sec:intro}

Interactive computer-based tasks generate considerably richer data than a single final response. In addition to whether a respondent solves a task, the digital environment records a time-stamped sequence of actions such as opening a page, selecting an option, entering information, or returning to an earlier step. For respondent $i$, these data may be viewed as a trajectory through a finite state space together with the waiting times between successive states. Thus, process data have both a \emph{sequential} component---which actions follow which---and a \emph{temporal} component---how quickly respondents move between them. Two respondents can arrive at the same final answer through different routes, or can follow the same route at very different speeds. These data therefore provide information about the organization of problem solving that is not available from response accuracy or total response time alone \citep{anghel2024use, he2025systematic}.

From a statistical perspective, such data raise several challenges. The state space can be moderately large, while only a small fraction of all possible transitions may be observed. Transition counts are therefore highly sparse and uneven, and the set of observed transitions may differ substantially across subpopulations. At the same time, both the timing and routing of transitions may vary systematically across respondents and groups. A useful analysis must therefore accommodate sparse event-history data, borrow information across related groups without eliminating meaningful heterogeneity, and characterize differences in entire transition structures rather than only in individual edges.

These issues arise naturally in large-scale international assessments. We consider process data from the Problem Solving in Technology-Rich Environments (PSTRE) component of the Programme for the International Assessment of Adult Competencies (PIAAC), which records detailed respondent interactions with computer-based tasks across multiple countries \citep{oecd2012literacy, ZA6712, oecd2019beyond}. Most cross-national analyses of assessment data focus on proficiency scores or response accuracy, whereas the processes leading to those outcomes have received substantially less attention \citep{ercikan2020use}. Existing work nevertheless suggests that countries with similar levels of performance can differ in the prevalence and timing of characteristic response patterns \citep{zheng2023identification}, and that the same digital behavior can differ across countries both in frequency and in its association with performance \citep{anghel2025highlighting}. This makes PIAAC process data a useful setting for studying heterogeneity in both transition dynamics and transition structure.

Several methodological traditions have been used to analyze process data. Sequence-mining methods identify frequently occurring or performance-related subsequences and solution strategies \citep{he2015identifying, he2021leveraging}, while probabilistic models describe action sequences and inter-action times through latent states or respondent-level traits \citep{chen2020continuous, xu2020latent, fu2024joint, tang2024latent}. A complementary representation treats actions as nodes and observed transitions as directed edges, producing transition networks that can be used to visualize solution paths, summarize behavioral structure, and identify subgroups \citep{zhu2016using, zhang2023identifying, saqr2025transition}. These approaches emphasize different features of the same stochastic process: event-history models describe transition intensities and waiting times, whereas network representations describe how probability is distributed among possible next actions. Integrating these two perspectives is particularly useful when both transition speed and global routing structure are of scientific interest.

Continuous-time models provide a natural framework for the temporal component. The same transition may occur quickly for one respondent and slowly for another, and the waiting time before a transition carries information about the pace of the response process. \citet{chen2020continuous}, for example, formulated process data as a marked point process in which the next action is modeled jointly with the waiting time. \citet{park2025analysis} instead adopted a multi-state survival formulation in which actions are states and the time between consecutive actions is a survival time. Transition-specific hazards can then incorporate respondent-level speed, covariate effects, and features of the source and destination actions. This representation is especially appealing because it separates the \emph{rate} at which a respondent leaves a state from the \emph{routing} among possible destination states.

Extending this framework to multiple countries introduces a hierarchical problem that cannot be addressed satisfactorily by fitting unrelated models country by country. In the PIAAC application, countries are further divided into correct- and incorrect-response groups, producing cells that vary substantially in size and contain many rare or unobserved transitions. Complete pooling would obscure country-level heterogeneity, whereas fitting separate models would yield unstable estimates in sparse cells and would not provide a direct estimate of between-country variation. Moreover, the observed transition support differs across country--response-group combinations, so the corresponding transition networks are not defined on identical edge sets. Finally, because the networks are functions of estimated transition parameters, uncertainty in the fitted model should be propagated into any subsequent analysis of network structure rather than discarded through a single plug-in estimate.

We develop a two-stage hierarchical framework for these data. In the first stage, we formulate a Bayesian multi-state survival model in which country-specific covariate effects, key-action effects, and country-level mean log-speed parameters are partially pooled across countries. Transition-specific baseline intensities are shared across countries within each response group, which stabilizes estimation for rare transitions, while hierarchical distributions allow the remaining effects to vary by country and quantify between-country heterogeneity. An interaction between source- and destination-key-action status allows transitions between two key actions to depart from an additive specification.

In the second stage, the fitted hazards are transformed into posterior transition probability matrices and treated as directed weighted networks. Because the observed supports differ across country--response-group combinations, elementwise comparison of transition matrices is not well suited to the problem. We instead use a modified directed graph auto-encoder \citep{kollias2022directed} to embed the networks in a common latent space with separate source and target representations for each action. Network dissimilarity is then quantified by the 1-Wasserstein distance between the resulting distributions of node-role embeddings \citep{togninalli2019wasserstein}. A common encoder is trained on posterior mean networks and then held fixed across posterior draws. Applying the encoder to each draw induces a posterior distribution of pairwise network dissimilarities, thereby propagating uncertainty from the multi-state model into the network analysis.

The resulting framework separates three related features of the response process. The multi-state model characterizes \emph{transition intensity}, including respondent characteristics, key-action effects, and overall speed. The hierarchical specification characterizes \emph{between-country heterogeneity} in these quantities. The network representation characterizes \emph{routing structure}, that is, how transition probability is distributed among possible next actions. This distinction is important because two groups may differ in how quickly they move through a task, in which transitions they favor, or in both.

The paper makes three main contributions. First, we develop a hierarchical multi-state survival model for sparse and unbalanced cross-national process data that jointly estimates population-level effects, country-specific effects, and between-country heterogeneity. Second, we develop a network representation of the model-implied transition probabilities that accommodates heterogeneous transition supports and propagates posterior uncertainty into measures of global network structure. Third, we use the combined framework to study response processes in two PIAAC problem-solving tasks across 14 countries, examining transition speed, covariate and key-action effects, global routing structure, and local differences in transition pathways.

The empirical analysis is organized around three questions: how respondent characteristics and key-action status are associated with transition intensities at the population and country levels; how much between-country heterogeneity is present in these associations and in transition speed; and how the global and local structure of transition networks varies across countries and response groups.

The remainder of the paper is organized as follows. Section~\ref{sec:data} describes the PIAAC process data, analytic samples, key actions, and transition sparsity. Section~\ref{sec:hmsm} presents the hierarchical multi-state survival model and the construction of posterior transition probability matrices. Section~\ref{sec:embed} introduces the directed graph embedding and Wasserstein-based network analysis. Section~\ref{sec:result} presents the empirical results, and Section~\ref{sec:discussion} discusses implications, limitations, and future directions.

\section{Data Description}\label{sec:data}

\subsection{Items, Sample, and Covariates}\label{sec:sample}

We analyze process data from the 2012 Problem Solving in Technology-Rich Environments (PSTRE) component of the Programme for the International Assessment of Adult Competencies (PIAAC), administered by the Organisation for Economic Co-operation and Development (OECD) \citep{oecd2012literacy,oecd2016survey}. The archived process-data files \citep{ZA6712} contain time-stamped records of respondents' interactions with the computer interface, including page changes, menu selections, mouse actions, and text entry. For each respondent and item, these records are converted into an ordered sequence of discrete actions together with the elapsed time between successive actions. Background characteristics are obtained from the corresponding PIAAC Public Use Files \citep{oecd2016survey}. We consider respondents from 14 countries and analyze two PSTRE items that differ substantially in their interface and task requirements.

The two items involve different problem-solving tasks and digital environments. In \emph{CD Tally} (U03A), respondents work in a combined web-browser and spreadsheet environment. The task places respondents in the role of an employee at a small music store whose manager has asked them to update the store's online inventory summary for the month. The spreadsheet lists information about multiple CDs and provides functions such as sorting and Find to help respondents locate and organize the records. Respondents must determine the number of Blues CDs, select the corresponding count in the online inventory form, and submit the form. Completing the task requires respondents to navigate between the two environments and use the spreadsheet information; common task-relevant actions include opening the data menu, using the sorting functions, sorting by genre, and returning to the website to select and submit the answer.

In \emph{Lamp Return} (U23X), respondents work in a combined email and web-browser environment. The task places respondents in the role of a customer who ordered a desk lamp online but received one in a different color from the color ordered. Respondents are asked to use the company's website to arrange an exchange for the lamp they originally ordered. To complete the task, they must navigate to the email environment, locate the relevant message, and retrieve the return authorization number. They then return to the website and navigate to the customer service and return pages. In the return form, respondents must indicate that the wrong item was shipped, request an exchange for the correct item, and enter the return authorization number obtained from the email before submitting the form. The task therefore requires respondents to coordinate information across the email and web environments while completing a sequence of related steps.

The two tasks consequently generate action spaces of different sizes and complexity. After preprocessing the log files, CD Tally contains $E=52$ distinct actions and Lamp Return contains $E=126$ distinct actions. These actions include both task-relevant operations and more general interface events, such as page navigation and keyboard input. A complete list and description of the distinct actions for each item, together with screenshots of the task interfaces, are provided in the Section~S1 of Supplementary Material.

The two items also differ in their scoring. CD Tally is scored dichotomously, so respondents are classified directly as having either a correct or an incorrect response. Lamp Return is scored on a scale from 0 to 3; we classify respondents receiving the maximum score of 3 as the \emph{correct-response group} and those receiving scores from 0 to 2 as the \emph{incorrect-response group}. We use these two response groups to examine whether the transition dynamics and routing patterns associated with successful task completion differ from those observed among respondents who do not fully complete the task. Together with the 14 countries, this classification yields 28 country--response-group combinations for each item.

Among the background variables available in PIAAC, we consider five respondent characteristics that capture demographic, socioeconomic, and technology-related differences that may be relevant to how respondents work through computer-based problem-solving tasks. These variables are gender, age, educational attainment, income, and \emph{Eskill}. Eskill is a standardized composite based on seven questions about the frequency of computer and Internet use and provides a measure of respondents' familiarity with digital technology. Education and income provide indicators of respondents' educational and socioeconomic backgrounds, while age and gender capture broader demographic differences. Details of the variable coding and descriptive statistics are provided in Section~S1.2 of the Supplementary Material.

\begin{table}[htb]
\centering
\caption{Sample sizes for the correct- and incorrect-response groups and proportions of correct responses by country and item. CD Tally is scored dichotomously. For Lamp Return, which is scored from 0 to 3, a score of 3 is classified as correct and scores of 0--2 as incorrect.}
\label{tab:sample}
\begin{tabular}{lccccccc}
\toprule
& \multicolumn{3}{c}{CD Tally} & & \multicolumn{3}{c}{Lamp Return} \\
\cmidrule(lr){2-4} \cmidrule(lr){6-8}
Country & Correct & Incorrect & Proportion correct & & Correct & Incorrect & Proportion correct \\
\midrule
Austria        & 437 & 217 & 0.67 & & 262 & 560 & 0.32 \\
Belgium        & 342 & 199 & 0.63 & & 304 & 443 & 0.41 \\
Germany        & 476 & 248 & 0.66 & & 318 & 656 & 0.33 \\
Denmark        & 491 & 361 & 0.58 & & 470 & 739 & 0.39 \\
Estonia        & 315 & 197 & 0.62 & & 303 & 411 & 0.42 \\
Finland        & 449 & 249 & 0.64 & & 499 & 432 & 0.54 \\
United Kingdom & 625 & 364 & 0.63 & & 534 & 817 & 0.40 \\
Ireland        & 298 & 189 & 0.61 & & 249 & 446 & 0.36 \\
South Korea    & 447 & 165 & 0.73 & & 316 & 528 & 0.37 \\
Netherlands    & 425 & 285 & 0.60 & & 457 & 559 & 0.45 \\
Norway         & 540 & 319 & 0.63 & & 429 & 623 & 0.41 \\
Poland         & 331 & 196 & 0.63 & & 341 & 487 & 0.41 \\
Slovakia       & 275 & 121 & 0.69 & & 222 & 370 & 0.38 \\
United States  & 335 & 221 & 0.60 & & 267 & 515 & 0.34 \\
\midrule
Total          & 5{,}786 & 3{,}331 & & & 4{,}971 & 7{,}586 & \\
\bottomrule
\end{tabular}
\end{table}

The analytic samples are restricted to respondents with complete information on these five covariates. This results in $N=9{,}117$ respondents for CD Tally and $N=12{,}557$ respondents for Lamp Return, with country-specific sample sizes reported in Table~\ref{tab:sample}. The proportion of respondents in the correct-response group ranges from $0.58$ to $0.73$ across countries for CD Tally and from $0.32$ to $0.54$ for Lamp Return. Consequently, the sample sizes of the 28 country--response-group combinations are unequal, with several cells containing fewer than 200 respondents. This imbalance, together with the sparsity of the observed transitions described in Section~\ref{sec:crossnational}, motivates the partial-pooling structure of the hierarchical model introduced in Section~\ref{sec:hmsm}.

\subsection{Key Actions}\label{sec:keyaction}

\begin{table}[htb!]
\centering
\small
\begin{tabular}{lll}
\toprule
Item & Action & Description \\
\midrule
\multirow{7}{*}{CD Tally}
& so\_1\_3 & Sort by third column (Genre) \\ 
& so\_2\_asc & Sort the spreadsheet in ascending order \\ 
& so\_2\_desc & Sort the spreadsheet in descending order \\ 
& so\_ok & Confirm sorting options \\ & so & Open the sort engine via the data menu \\ 
& ss\_data & Click the data menu on the spreadsheet page \\ 
& ss\_so & Open the sort engine on the spreadsheet page \\
\midrule
\multirow{13}{*}{Lamp Return}
& wb\_pg\_8\_4\_reason\_4 & Select the reason for return (wrong item) \\ 
& wb\_pg\_8\_4\_submit & Submit the return form \\ 
& wb\_pg\_8\_4\_request\_1 & Select the request (exchange) \\ 
& wb\_pg\_8\_4 & Open the return form \\ 
& wb\_pg\_8 & Link to the Customer Service page \\ 
& wb\_pg\_8\_2 & View updated orders and shipping \\ 
& wb\_pg\_8\_3 & Obtain the authorization number \\ 
& wb & Switch to the website page \\ 
& wb\_hist\_back & Go to the previous page \\ 
& em & Switch to the email page \\ 
& em\_m\_view\_305 & View email 305 (authorization number) \\ 
& em\_f\_view & View the email folder \\ 
& keypress & Press a keyboard key \\
\bottomrule
\end{tabular}
\caption{Key actions identified for the CD Tally and Lamp Return items. Key actions are those whose occurrence is most strongly associated with response correctness and are used to characterize transitions involving task-relevant actions.}
\label{tab:keyactions}
\end{table}

Because the action spaces are relatively large, it is useful to distinguish a smaller subset of actions that are especially informative about successful task completion. We refer to these as \emph{key actions}. Identifying such actions serves two purposes. First, it provides a substantively interpretable summary of the parts of the response process that are most strongly associated with response correctness. Second, it allows the multi-state survival model to capture systematic differences in transition intensity involving task-relevant actions without introducing separate regression effects for every individual action.

We define the set of key actions, denoted by $\mathcal A$, as those whose occurrence is most strongly associated with whether a respondent answers the item correctly. The key-action sets are identified using the $\chi^2$ procedure of \citet{he2015identifying}. For each action, the procedure measures the association between its weighted occurrence frequency and response correctness; actions are then ranked by this association, and those above an elbow point in the ordered scores are retained \citep{park2025analysis}.

Table~\ref{tab:keyactions} lists the resulting seven key actions for CD Tally and thirteen for Lamp Return. For CD Tally, the selected actions primarily involve spreadsheet sorting operations that are central to identifying the requested CD genre. For Lamp Return, they primarily involve navigating the return process, obtaining the authorization number, and completing the online return form. The complete action inventories, including descriptions of both key and non-key actions, are provided in Section~S1.4 of the Supplementary Material.

\subsection{Cross-National Structure and Transition-Level Sparsity}\label{sec:crossnational}

Let $g=(q,c)$ denote a combination of country $q=1,\ldots,Q$ ($Q=14$) and response group $c\in\{0,1\}$ with $c=1$ for correct responses (hereafter a country-group combination) and let the observed transition support of $g$ be the set of ordered action pairs $(m,l)$ with $m\neq l$ observed at least once among respondents in $g$. There are $G=2Q=28$ combinations per item. Three features of these data motivate the modeling strategy. First, the data are nested and the cells are small and unbalanced. As shown in Table~\ref{tab:sample}, cell sizes range from 121 to 625 for CD Tally and from 222 to 817 for Lamp Return whereas the countries differ in the proportion of correct responses. A single set of coefficients would suppress differences that the data indicate whereas separate coefficients for every country would rest on as few as 121 respondents. A hierarchical specification estimates country-specific coefficients and their common distribution so that the degree of shrinkage is determined by the data \citep{gelman2007data}.

\begin{table}[htb!]
\centering
\caption{Structural summary of the observed transition data by item and response group. Entries are minimum / median / maximum across the 14 countries in each row. Support size is the number of ordered action pairs observed at least once. Coverage is the support size as a percentage of the $E(E-1)$ possible pairs. Pairs $<5$ is the percentage of observed pairs occurring fewer than five times. 
% Jaccard is the pairwise Jaccard similarity between the supports of all 28 combinations of an item (minimum / median / maximum over pairs).
}
\label{tab:structure}
\begin{tabular}{llcccc}
\toprule
Item & Group & $n$ & Support size & Coverage (\%) & Pairs $<5$ (\%) \\
\midrule
CD Tally    & Correct   & 275 / 431 / 625 & 205 / 422 / 483 & 7.7 / 15.9 / 18.2 & 66.7 / 73.0 / 77.6 \\ & Incorrect & 121 / 219 / 364 & 142 / 302 / 397 & 5.4 / 11.4 / 15.0 & 72.2 / 82.0 / 92.3 \\
\addlinespace
Lamp Return & Correct   & 222 / 317 / 534 & 354 / 513 / 717 & 2.2 / 3.3 / 4.6   & 63.5 / 67.6 / 73.2 \\ & Incorrect & 370 / 522 / 817 & 494 / 695 / 998 & 3.1 / 4.4 / 6.3   & 67.5 / 70.3 / 77.0 \\
\bottomrule
\end{tabular}
\end{table}

Second, the supports are heterogeneous in size and composition. As shown in Table~\ref{tab:structure} support sizes range from 142 to 483 of the $E(E-1) = 2{,}652$ possible pairs for CD Tally and from 354 to 998 of $15{,}750$ for Lamp Return. The pairwise Jaccard similarity between supports has median 0.46 for CD Tally and 0.45 for Lamp Return and the overlap coefficient which adjusts for differences in support size has median 0.73 and 0.71. Country-group combinations thus traverse partly distinct regions of the action space and a comparison of transition structures must accommodate networks defined on different supports. Third, transition counts are concentrated on rare pairs. In all 28 combinations at least half of the observed pairs occur three times or fewer and the proportion occurring fewer than five times ranges from 63.5\% to 92.3\% as reported in the last column of Table~\ref{tab:structure}. Separate estimation of transition-specific baselines within each combination would therefore rest on little information for most pairs. The baseline intensities are hence shared across countries within each response group and country-level variation is carried by the covariate effects, the key-action effects, and the country-specific speed distributions.

\section{Hierarchical Multi-state Survival Models}\label{sec:hmsm}

\subsection{Model Specification}\label{sec:model}

A multi-state survival model \citep[MSM;][]{commenges1999multi, hougaard1999multi, andersen2002competing, putter2007tutorial, meira2009multi, crowther2017parametric} represents an action sequence as a trajectory through a finite set of states in which the time between consecutive actions is a survival time and transition intensities are hazard functions. We extend the Bayesian multi-state survival model \citep{park2025analysis} in two main ways. First, whereas their model was fitted separately to each country, we introduce a hierarchical structure across countries, enabling partial pooling and estimation of between-country heterogeneity. Second, we add an interaction between source- and destination-key-action status, allowing the key-action contribution to the log-hazard of key-to-key transitions to depart from additivity.

We retain the time-homogeneous Markov assumption under which transition rates are constant over time and depend only on the current state. We define the model on a common transition support $\mathcal T$, the set of off-diagonal transitions observed in the pooled sample. For $(m,l)\in\mathcal T$, the hazard of the transition from action $m$ to action $l$ for respondent $i$ in country $q$ is
\begin{equation}\label{eq:hierarchical_hazard}
    \lambda_{m,l,i,q} = \kappa_{c_i,m,l} \, \tau_i \exp\!\Biggl\{ \sum_{p=1}^{P} \alpha_{p,q}\, x_{i,p} + \beta_{c_i,1,q}\, I(m \in \mathcal A) + \beta_{c_i,2,q}\, I(l \in \mathcal A) + \beta_{c_i,3,q} I(m \in \mathcal A) I(l \in \mathcal A) \Biggr\},
\end{equation}
where $P$ is the number of covariates and $\mathcal A$ is the set of key actions. The baseline hazard $\kappa_{c_i,m,l}>0$ is specific to response group $c_i\in\{0,1\}$ and shared across countries. The speed parameter $\tau_i>0$ captures the overall pace of respondent $i$. The coefficient $\alpha_{p,q}$ is the effect of covariate $p$ on the log-hazard in country $q$. The parameters $\beta_{c_i,1,q}$ and $\beta_{c_i,2,q}$ are the source-key and destination-key main effects and $\beta_{c_i,3,q}$ is their interaction which gives the additional log-hazard when both source and destination are key actions. 

The country-specific effects are partially pooled. For each covariate $p$ the coefficients $\{\alpha_{p,q}\}_{q=1}^{Q}$ share a population mean $\mu_{\alpha,p}$ and a between-country variance $\sigma_{\alpha,p}^{2}$. For each response group $c$ and effect type $r$ the coefficients $\{\beta_{c,r,q}\}_{q=1}^{Q}$ share $\mu_{\beta,c,r}$ and $\sigma_{\beta,c,r}^{2}$. The parameters of the country-specific distributions of $\tau_i$ follow a parallel hierarchy. The variance components quantify between-country heterogeneity and the partial pooling stabilizes estimation in small cells.

\subsection{Prior and Computation}\label{sec:prior}

Let $a_{i,j}$ denote the $j$th action of respondent $i$ for $j=0, 1, \cdots, E_i$ where $a_{i,0}$ is the initial action and $E_i$ is the number of observed transitions. Let $t_{i,j}$ denote the time of the $j$th transition with $t_{i,0}=0$ and let $q(i)$ denote the country of respondent $i$. Define the transition count $D_{m,l,i} = \sum_{j=1}^{E_i} I(a_{i,j-1} = m, a_{i,j} = l)$ and the time at risk $R_{m,i} = \sum_{j=1}^{E_i} (t_{i,j} - t_{i,j-1}) I(a_{i,j-1} = m)$. Under the time-homogeneous Markov assumption the likelihood of the data $\mathcal Y$ given the parameters $\Theta$ is
\begin{equation}\label{eq:likelihood}
L(\mathcal Y\mid \boldsymbol\Theta) = \prod_{i=1}^{N} \prod_{m=1}^{E} \prod_{(m,l)\in\mathcal T} \lambda_{m,l,i,q(i)}^{D_{m,l,i}} \exp \left(-\lambda_{m,l,i,q(i)} R_{m,i}\right).
\end{equation}

Let $r=1,2,3$ index the source-key main effect, the destination-key main effect, and the interaction. The priors are
\begin{equation}\begin{split}
    \label{eq:hierarchical_priors}
    \kappa_{c,m,l} &\sim \mathrm{LN}(\mu_\kappa,\sigma_\kappa^{2}), \quad 
    \tau_i \sim \mathrm{LN}(\mu_{\tau,q(i)},\sigma_{\tau,q(i)}^{2}), \quad 
    \sigma_{\tau,q}^{2} \sim \mathrm{IG}(a_\tau,b_\tau), \quad
    \mu_{\tau,q} \sim \mathrm{N}(0,\sigma_{\mu_\tau}^{2}), \\
    \sigma_{\mu_\tau}^{2} &\sim \mathrm{IG}(c_\tau,d_\tau), \quad
    \alpha_{p,q} \sim \mathrm{N}(\mu_{\alpha,p}, \sigma_{\alpha,p}^{2}), \quad
    \mu_{\alpha,p} \sim \mathrm{N}(0,\sigma_{\alpha}^{2}), \quad
    \sigma_{\alpha,p}^{2} \sim \mathrm{IG}(a_\alpha,b_\alpha), \\
    \beta_{c,r,q} &\sim \mathrm{N}(\mu_{\beta,c,r},\sigma_{\beta,c,r}^{2}), \quad
    \mu_{\beta,c,r} \sim \mathrm{N}(0,\sigma_{\mu_\beta}^{2}), \quad
    \sigma_{\beta,c,r}^{2}\sim\mathrm{IG}(a_\beta,b_\beta),    
\end{split}\end{equation}
where LN and IG denote the log-normal and inverse-gamma distributions and the indices run over $c \in \{0,1\}$, $(m,l) \in \mathcal T$, $i=1, \cdots, N$, $q=1, \cdots, Q$, $p = 1, \cdots, P$, and $r=1,2,3$. The posterior distribution is
\begin{equation}\begin{split}
\label{eq:hierarchical_posterior}
\pi\Big(\Theta \mid \mathcal Y\Big) &\propto L\Big(\mathcal Y\mid\Theta\Big) \prod_{c,m,(m,l)\in \mathcal T} \pi\Big(\kappa_{c,m,l}\Big) \prod_{i} \pi\Big(\tau_i\mid\mu_{\tau,q(i)}, \sigma_{\tau,q(i)}^{2}\Big) \\
&\times \prod_{q} \pi\Big(\mu_{\tau,q} \mid \sigma_{\mu_\tau}^{2}\Big) \pi\Big(\sigma_{\tau,q}^{2}\Big) \pi\Big(\sigma_{\mu_\tau}^{2}\Big) \prod_{p,q} \pi\Big(\alpha_{p,q}\mid\mu_{\alpha,p}, \sigma_{\alpha,p}^{2}\Big) \prod_{p} \pi\Big(\mu_{\alpha,p}\Big) \pi\Big(\sigma_{\alpha,p}^{2}\Big)\\
&\times \prod_{c,r,q} \pi\Big(\beta_{c,r,q}\mid\mu_{\beta,c,r}, \sigma_{\beta,c,r}^{2}\Big) \prod_{c,r} \pi\Big(\mu_{\beta,c,r}\Big) \pi\Big(\sigma_{\beta,c,r}^{2}\Big),
\end{split}\end{equation}
where $L(\mathcal Y\mid\Theta)$ is the likelihood in \eqref{eq:likelihood} and each prior density is given in \eqref{eq:hierarchical_priors}. The variances $\sigma_\alpha^2$ and $\sigma_{\mu_\beta}^2$ of the population means are fixed whereas $\sigma_{\alpha,p}^2$ and $\sigma_{\beta,c,r}^2$ are estimated between-country variances. We set $\mu_\kappa=0$, $\sigma_\kappa^2 = \sigma_\alpha^2 = \sigma_{\mu_\beta}^2=1$, and all inverse-gamma shape and scale parameters to $0.001$.

Posterior inference used four MCMC chains of 300,000 iterations each. The first 100,000 iterations were discarded and every 20th draw was retained so that each chain yielded 10,000 draws. Proposal variances of the Metropolis--Hastings updates were tuned to acceptance rates between 0.2 and 0.5. Convergence was assessed with the rank-normalized split-$\widehat R$ diagnostic \citep{vehtari2021rank}. Implementation details are given in the Section~S2 of the Supplementary Material.

\subsection{Transition Probability Matrices}\label{sec:matrix}

The fitted hazards are converted into transition probability matrices that give the conditional distribution of the next action given the current action. Let $\mathcal S=\{1,\ldots,E\}$ denote the set of all actions of an item and let $g(i)=(q(i),c_i)$ denote the country-group combination of respondent $i$. For combination $g$, let $N_g(m,l)=\sum_{i:g(i)=g}D_{m,l,i}$ denote the number of observed transitions from $m$ to $l$ and $n_g(m) = \sum_{l\neq m}N_g(m,l)$ the number of observed transitions out of $m$. The empirical transition support is $\mathcal E_g=\{(m,l): m\neq l,\; N_g(m,l)>0,\; n_g(m)\geq\nu_0\}$ with $\nu_0=3$ so that transition probabilities are constructed only for source actions with at least three observed exits. Let $\mathcal N_g^+(m)=\{u\in\mathcal S:(m,u)\in\mathcal E_g\}$ denote the observed destinations of $m$.

Row-normalizing the hazards in \eqref{eq:hierarchical_hazard} over $\mathcal N_g^+(m)$ at posterior draw $s$ gives for $g = (q, c)$
\begin{equation}\label{eq:transition_prob}
A^{(s,g)}_{m,l}=\frac{\kappa^{(s)}_{c,m,l}\exp\{\eta^{(s)}_{c,q}(m,l)\}}{\sum_{u\in\mathcal N_g^+(m)}\kappa^{(s)}_{c,m,u}\exp\{\eta^{(s)}_{c,q}(m,u)\}} \quad \text{for} \quad l\in\mathcal N_g^+(m),
\end{equation}
where $\eta^{(s)}_{c,q}(m,l) = \Big\{\beta^{(s)}_{c,2,q} + \beta^{(s)}_{c,3,q} I(m\in\mathcal A) \Big\} I(l\in\mathcal A)$ and $A^{(s,g)}_{m,l}=0$ otherwise. The speed parameter $\tau_i$, the covariate effects $\alpha_{p,q}$, and the source-key effect $\beta_{c,1,q}$ multiply all outgoing hazards of a source action by a common factor. They govern the exit rate and the time spent in the current action and cancel under the normalization so that $A^{(s,g)}$ does not depend on respondent-level quantities and no reference covariate profile is needed. The baseline hazards $\kappa_{c,m,l}$ and the routing effects $\beta_{c,2,q}$ and $\beta_{c,3,q}$ vary across destinations and determine how probability is allocated among the outgoing edges. Because $\kappa_{c,m,l}$ is shared across countries within each response group cross-country differences in the networks arise only through $(\beta_{c,2,q}, \beta_{c,3,q})$ and the supports $\mathcal E_g$. Each nonzero row of $A^{(s,g)}$ sums to one over the observed destinations. Rows with $\mathcal N_g^+(m)=\emptyset$ are zero and self-transitions are excluded. The entry $A^{(s,g)}_{m,l}$ is the embedded probability of the next transition rather than a finite-time transition probability. 

\section{Embedding and Comparison of Transition Probability Networks}\label{sec:embed}

The transition probability matrix $A^{(s,g)}$ of each country-group combination $g$ at posterior draw $s$ is a directed weighted network on the support $\mathcal E_g$. These matrices are deterministic functions of the model parameters rather than parameters themselves and the posterior distribution of the parameters therefore induces a posterior distribution over the network of every combination. The goal of this section is to compare the networks across combinations and to identify structural similarities and differences among countries and response groups.

Two features of the networks complicate this comparison. The supports $\mathcal E_g$ differ in size and composition so that a source action without observed exits in one combination may be well populated in another. Elementwise distances between transition matrices treat such support differences as differences in probability values and ignore the relational structure of directed graphs. We therefore proceed in two stages. A modified directed graph auto-encoder (DiGAE) adapted to row-stochastic matrices with heterogeneous supports first embeds all networks in a common latent space. The 1-Wasserstein distance between the node embedding distributions of two networks is then computed for each posterior draw so that posterior uncertainty in the networks is carried into the pairwise distances. Averaging the draw-specific distances over the posterior sample yields a distance matrix that summarizes the dissimilarity among the 28 networks of an item.

\subsection{Directed Graph Auto-Encoder (DiGAE) Framework}\label{sec:digae}

The transition networks are directed and weighted and the direction of an edge is substantively meaningful. We therefore use the directed graph auto-encoder \citep[DiGAE;][]{kollias2022directed} which learns two latent vectors for each node. Let $d$ denote the embedding dimension. Each action $m$ has a source embedding $z_{S,m}\in\mathbb R^d$ that encodes its role as the origin of transitions and a target embedding $z_{T,m}\in\mathbb R^d$ that encodes its role as the destination of transitions. The encoder consists of directed graph convolutional layers that update source embeddings by aggregating over target neighbors and target embeddings by aggregating over source neighbors. Collecting the embeddings as rows of $Z_S$ and $Z_T\in\mathbb R^{E\times d}$ the decoder reconstructs the network as $\bar A=\varsigma(Z_S Z_T^\top)$ where $\varsigma$ is the elementwise sigmoid function. The asymmetric inner product $z_{S,m}^\top z_{T,l}$ allows the edges $m\to l$ and $l\to m$ to receive different weights.

Three properties of the transition networks motivate the modifications in Section~\ref{sec:adaptation}. Each nonzero row of $A^{(s,g)}$ is a probability distribution without self-transitions whereas the standard decoder predicts independent edge probabilities. The node and edge supports differ across combinations. The trained encoder is applied to all posterior draws rather than to a single network so that posterior uncertainty is carried into the embedding space.

\subsection{Adaptations for Transition Probability Networks}\label{sec:adaptation}

\subsubsection{Graph Convolutional Encoder}\label{sec:encoder}

We retain the encoder \citep{kollias2022directed} and modify the treatment of the adjacency matrix. Let $\tilde A$ denote the augmented adjacency matrix defined below and let $\tilde D^+$ and $\tilde D^-$ denote the diagonal matrices of its out-degrees and in-degrees. With $\hat A=(\tilde D^+)^{-\gamma^+}\tilde A(\tilde D^-)^{-\gamma^-}$, the updates at layer $\ell$ are 
\[
    S^{(\ell+1)}=\hat A\,T^{(\ell)}W_T^{(\ell)} \quad \mbox{and} \quad T^{(\ell+1)}=\hat A^\top S^{(\ell)}W_S^{(\ell)},
\]
where $W_S^{(\ell)}$ and $W_T^{(\ell)}$ are trainable weight matrices and $\gamma^+,\gamma^-\in[0,1]$ control the strength of out-degree and in-degree normalization. 

Standard DiGAE sets $\tilde A=A+I$ and adds a self-loop to every node. We instead distinguish nodes with at least one outgoing transition, sink nodes with incoming transitions only, and isolated nodes with no transitions and define
\begin{equation}\label{eq:augmented}
\begin{array}{ll}
     \tilde A_{ml} = A_{ml} & \text{if } \textstyle\sum_k A_{mk}>0, \\
     \tilde A_{ml} = \delta_{ml} & \text{if } \textstyle\sum_k A_{mk}=0 \text{ and } \sum_k A_{km}>0, \\
     \tilde A_{ml} = 0 & \text{otherwise},
\end{array}
\end{equation}
where $\delta_{ml}$ is the Kronecker delta. A node with outgoing transitions keeps its row of $A$ so that the transition structure is preserved. A sink node receives a self-loop so that its representation remains informative across the layers. An isolated node retains a zero row and a zero column. Sink and isolated nodes have empty outgoing supports and are excluded from the reconstruction loss below.

We use two layers with ReLU nonlinearity and the identity matrix $X=I_E$ as node features so that each action enters as a one-hot vector. The encoder is
\[
    Z_S=\hat A\,\mathrm{ReLU}(\hat A^\top X W_S^{(0)})W_T^{(1)} \quad \mbox{and} \quad Z_T=\hat A^\top\mathrm{ReLU}(\hat A X W_T^{(0)})W_S^{(1)}
\]
where $W_S^{(0)},W_T^{(0)}\in\mathbb R^{E\times d_1}$ and $W_S^{(1)},W_T^{(1)} \in \mathbb R^{d_1\times d}$ where $d_1$ is the hidden dimension. The cross use of $W_T^{(1)}$ in the source path and $W_S^{(1)}$ in the target path follows the block-antidiagonal parameterization \citep{kollias2022directed} so that the source and target embeddings are learned jointly rather than as two separate encoders. % For the Wasserstein comparison in Section~\ref{sec:wasserstein} each action $m$ is represented by the concatenation $z_m=[z_{S,m};z_{T,m}]\in\mathbb R^{2d}$.

\subsubsection{Decoder and Loss Function}\label{sec:decoder}

The standard decoder $\bar A=\varsigma(Z_SZ_T^\top)$ predicts an independent existence probability for each edge. This decoder does not match the transition networks because each nonzero row of $A^{(g)}$ is a probability distribution over the observed destinations $\mathcal N_g^+(m)$ and self-transitions are excluded by construction. Let $\mathcal M_g=\{m\in\mathcal S:\mathcal N_g^+(m)\neq\emptyset\}$ denote the source actions with at least one observed destination and let $\Psi^{(g)}$ denote the mask with $\Psi^{(g)}_{ml}=0$ for $(m,l)\in\mathcal E_g$ and $\Psi^{(g)}_{ml}=-\infty$ otherwise under the convention $\exp(-\infty)=0$. The decoder is
\begin{equation}\label{eq:decoder}
\bar A^{(g)} = \mathrm{Softmax}_{\mathrm{row}} \Big(Z_SZ_T^\top+\Psi^{(g)}\Big),
\end{equation}
where the softmax is applied to each row $m\in\mathcal M_g$ and rows with $m\notin\mathcal M_g$ are set to zero. The decoder allocates probability only among the observed destinations so that $\bar A^{(g)}$ has the support of $A^{(g)}$ and $\bar A^{(g)}_{ml}$ is the reconstructed conditional probability of action $l$ given action $m$. The diagonal is masked because $\mathcal E_g$ excludes self-transitions and the self-loops added to sink nodes in the encoder therefore affect message passing only and do not enter the reconstructed network.

The encoder is trained on the posterior mean networks $A^{(g)}=N_{\mathrm{draw}}^{-1}\sum_{s}A^{(s,g)}$ by minimizing the forward Kullback--Leibler divergence averaged over source actions with observed exits,
\begin{equation}\label{eq:loss}
\mathcal L^{(g)}=|\mathcal M_g|^{-1}\sum_{m\in\mathcal M_g}\mathrm{KL}\bigl(A^{(g)}_{m:}\,\|\,\bar A^{(g)}_{m:}\bigr),
\end{equation}
with the convention $0\log(0/q)=0$. A single encoder is trained jointly on the $G$ posterior mean networks with the Adam optimizer \citep{kingma2015iclr-adam} and is held fixed when applied to the posterior draws. Auto-encoder representations are identifiable only up to transformations of the latent coordinates so that separate encoders per draw would yield latent spaces without a common scale. A fixed encoder places all networks and all draws in one latent space in which distances are comparable across graphs and across draws. Posterior uncertainty is thus propagated through the encoder inputs rather than through the encoder parameters. Reconstruction losses on the posterior draws are comparable to those on the training networks and are provided in the Section~S3.3 of the Supplementary Material.

\subsection{Wasserstein Distance-based Network Comparison}\label{sec:wasserstein}

The fixed encoder maps each network $A^{(s,g)}$ to a set of node embeddings $Z^{(s,g)}=\{z^{(s,g)}_m\}\subset\mathbb R^{2d}$ in a common latent space where $z_m=[z_{S,m};z_{T,m}]$ concatenates the source and target embeddings of action $m$. Comparing two networks then amounts to comparing two discrete distributions over $\mathbb R^{2d}$. We use the 1-Wasserstein distance for two reasons. First, the supports differ across combinations so that the sets of actions with positive mass differ. Optimal transport moves mass between nearby embeddings and aligns structurally similar node roles without forcing the same action to be matched across networks. The distance is therefore a dissimilarity between distributions of learned node roles rather than an action-by-action discrepancy. Second, aggregating node embeddings by summation or averaging discards distributional information that may carry differences between graphs \citep{togninalli2019wasserstein}. Following \citet{togninalli2019wasserstein}, we use the Euclidean ground cost which places less emphasis on large embedding differences than squared-cost alternatives such as $W_2$. Section~S4 of the Supplementary Material reports a sensitivity analysis in which the pairwise distances were also computed with the energy distance \citep{szekely2013energy}, maximum mean discrepancy with an RBF kernel \citep{gretton2012kernel}, and the Chamfer distance \citep{tenenbaum1977parametric, fan2017point} and compared with the Wasserstein distances through Spearman rank correlations of the resulting distance matrices.

For graphs $g$ and $h$ at draw $s$ let $C^{(s;g,h)}_{ml}=\|z^{(s,g)}_m-z^{(s,h)}_l\|_2$ and let $a$ and $b$ denote uniform distributions over the nodes that participate in at least one transition of the respective graph with zero mass on isolated nodes. The draw-specific distance is the optimal transport cost \citep{peyre2020computationaloptimaltransport}
\begin{equation}\label{eq:wasserstein}
d^{(s)}(g,h)=\min_{\Gamma\in U(a,b)}\langle\Gamma,C^{(s;g,h)}\rangle,
\end{equation}
where $U(a,b)$ is the set of nonnegative matrices with row sums $a$ and column sums $b$ and $\langle\Gamma,C\rangle=\sum_{m,l}\Gamma_{ml}C_{ml}$. Isolated nodes have zero rows and columns in the augmented adjacency matrix and carry no transition information so that excluding them keeps the distance focused on the roles of participating actions. The same draw $s$ is used for both graphs so that $d^{(s)}(g,h)$ reflects a joint draw of the hierarchical model and preserves the posterior dependence between the two networks. Conditional on the fixed encoder the collection $\{d^{(s)}(g,h)\}_{s=1}^{N_{\mathrm{draw}}}$ is an induced posterior distribution of the dissimilarity between $g$ and $h$. We summarize it by the posterior mean $\bar D_{gh}=N_{\mathrm{draw}}^{-1}\sum_{s}d^{(s)}(g,h)$ and by the 95\% highest posterior density (HPD) interval. % The $G\times G$ matrix $\bar D$ is the input to nMDS, PERMANOVA, and PERMDISP.

\section{Results}\label{sec:result}

% We applied our analysis framework to CD Tally and Lamp Return item in PSTRE. For each item we first summarize the posterior estimates of the hierarchical survival model (country-level speed, covariate effects, key-action effects, and baseline intensities) and then compare the posterior transition networks through the DiGAE embedding, Wasserstein distances, nMDS, PERMANOVA, PERMDISP, and an additive and multiplicative effects (AME) decomposition of selected contrasts. The maximum rank-normalized split-$\widehat R$ across parameters was 1.03 for CD Tally and 1.06 for Lamp Return. Complete diagnostics and trace plots are given in the Supplementary Material and in the accompanying code repository.

We applied the proposed framework to the CD Tally and Lamp Return items of the PSTRE assessment in PIAAC. For each item we first summarize the posterior estimates of the hierarchical MSM and then compare the posterior transition networks in two steps. Distance-based analyses of the Wasserstein distance matrix characterize the global structure of the 28 networks and a decomposition of selected pairwise contrasts examines local differences in transition structure. The maximum rank-normalized split-$\widehat R$ across parameters was 1.03 for CD Tally and 1.06 for Lamp Return. Complete diagnostics and trace plots are given in the Section~S2.5 of the Supplementary Material and in the accompanying code repository.

\subsection{CD Tally Item}\label{sec:cdtally}

\subsubsection{Parameter Estimation}\label{sec:cdpara}

\begin{figure}[htb!]
    \centering
    \includegraphics[width=0.75\linewidth]{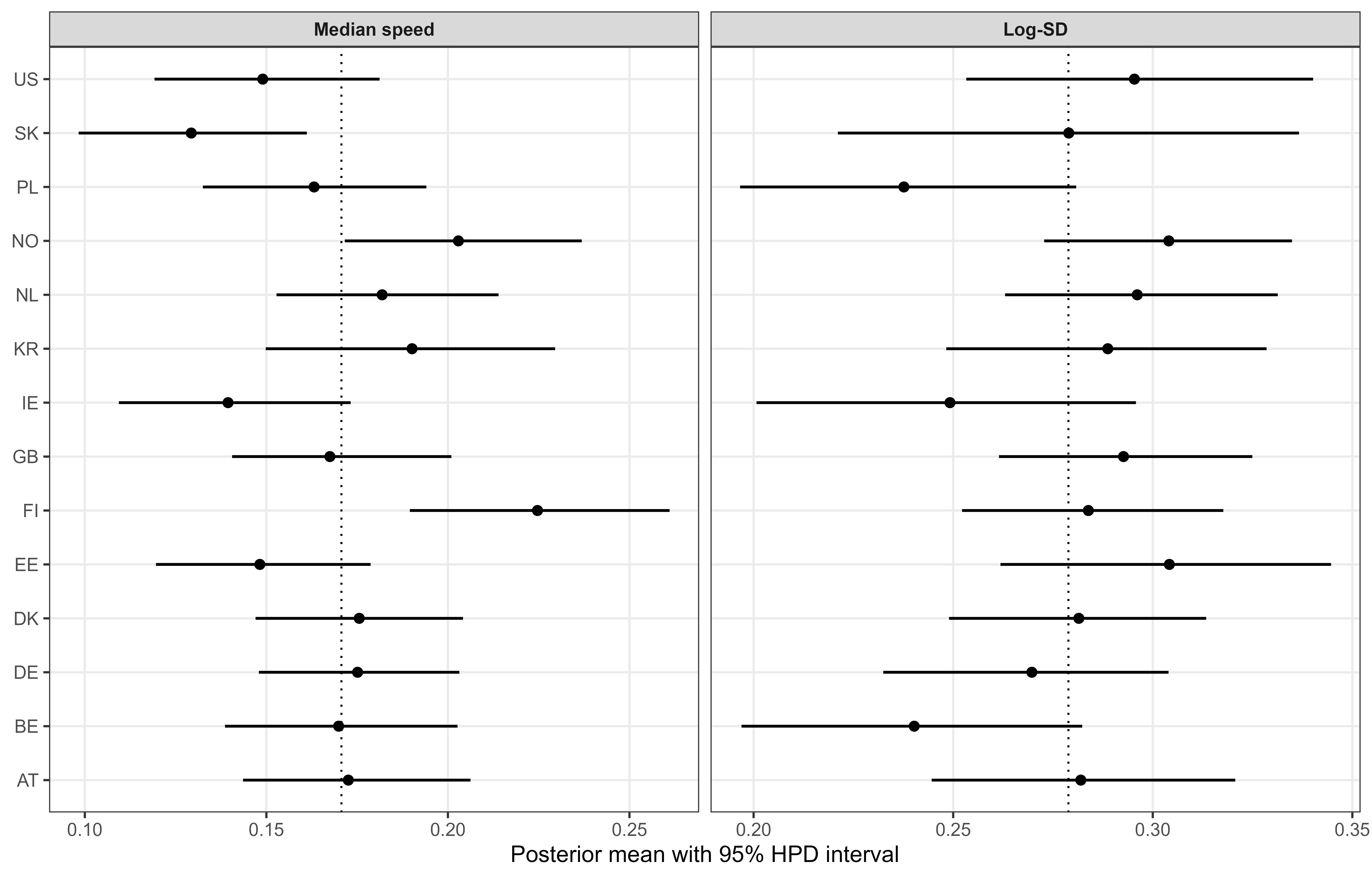}
    \caption{Posterior means and 95\% HPD intervals of the country-level median speed $\exp(\mu_{\tau,q})$ (left) and the within-country standard deviation of log speed $\sigma_{\tau,q}$ (right) for CD Tally. Dotted lines mark the means across countries.}
    \label{fig:tau}
\end{figure}

\paragraph{Country-level Speed} Figure~\ref{fig:tau} shows the posterior means of the country-level median speed $\exp(\mu_{\tau,q})$ and of the within-country standard deviation of log speed $\sigma_{\tau,q}$. Countries are labeled by their two-letter codes throughout the results and the correspondence between codes and country names is given in the Section~S1.1 of the Supplementary Material. The median speed varies across countries. Finland (FI) is fastest and Slovakia (SK) slowest and their 95\% HPD intervals do not overlap. FI, NO, KR, NL, and DK lie at the upper end and SK, IE, EE, and US at the lower end. The within-country heterogeneity in log speed is roughly similar across countries.

\begin{figure}[htb!]
    \centering
    \includegraphics[width=1\linewidth]{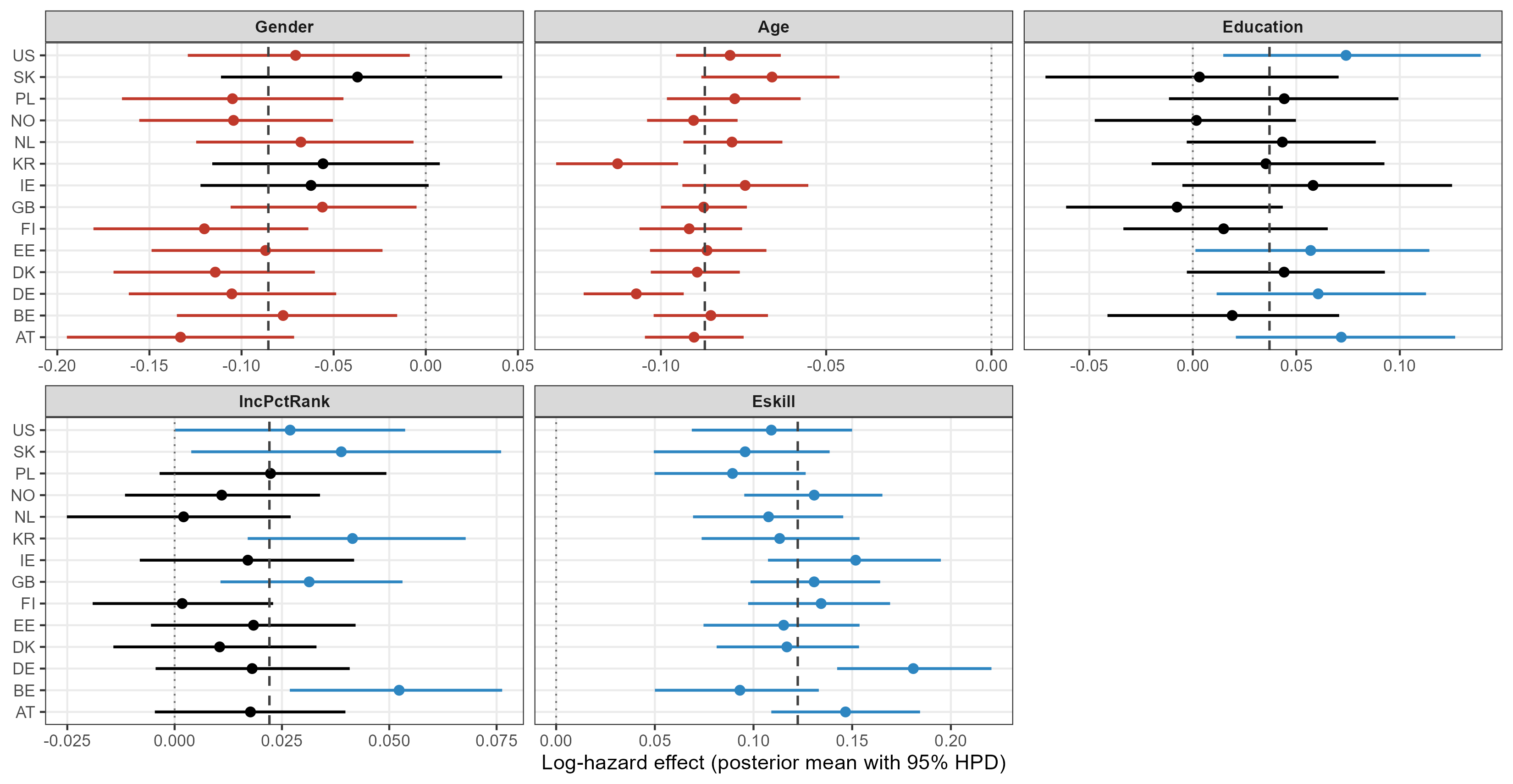}
    \caption{Country-specific covariate effects $\alpha_{p,q}$ on the log-hazard for CD Tally. Points are posterior means and bars are 95\% HPD intervals. Blue and red mark intervals above and below zero and black marks intervals that include zero. Dashed lines mark the pooled effects $\mu_{\alpha,p}$ and dotted lines mark zero.}
    \label{fig:alpha_cd}
\end{figure}

\paragraph{Covariate Effects} Figure~\ref{fig:alpha_cd} shows the country-specific covariate effects $\alpha_{p,q}$. Age is credibly negative and Eskill credibly positive in all 14 countries. Gender is negative in all countries and credibly negative in 11 of them. Education is mostly positive and credibly positive in four countries. Income has the smallest effects and is credibly positive in five countries. Age and Eskill thus show the most consistent cross-national associations with transition intensity.

\paragraph{Key-action Effects} Table~\ref{tab:key_action_effects_cd} reports the country-specific key-action effects. Relative to a transition between two non-key actions the log-hazard changes by $\beta_{c,1,q}$ for a transition from a key action to a non-key action, by $\beta_{c,2,q}$ for a transition from a non-key action to a key action, and by $\beta_{c,1,q}+\beta_{c,2,q}+\beta_{c,3,q}$ for a transition between two key actions. In the correct group $\beta_{1,1,q}$ is credibly negative and $\beta_{1,2,q}$, $\beta_{1,3,q}$, and the total effect are credibly positive in all 14 countries. Correct respondents thus move from key actions to non-key actions at lower intensity and move into and among key actions at higher intensity. In the incorrect group neither main effect has an interval excluding zero in any country whereas $\beta_{0,3,q}$ and the total effect are credibly positive in all countries. The contrasts $\Delta\beta_{1,q}=\beta_{1,1,q}-\beta_{0,1,q}$ and $\Delta\beta_{2,q}$ are credibly negative and credibly positive in all countries whereas the intervals of $\Delta\beta_{3,q}$ and of the total contrast include zero in every country. The response groups therefore differ at the boundary of the key-action set rather than in the elevation of key-to-key transitions which is present in both groups.

\begin{table}[htb!]
\centering
\caption{Posterior means of key-action effects by country for CD Tally. The total effect $\sum_{r=1}^{3}\beta_{c,r,q}$ is the log-hazard effect of a key-to-key transition. The contrast is $\Delta\beta_{r,q}=\beta_{1,r,q}-\beta_{0,r,q}$. Blue and red mark 95\% HPD intervals above and below zero and black marks intervals that include zero.}
\label{tab:key_action_effects_cd}
\centering
\resizebox{\ifdim\width>\linewidth\linewidth\else\width\fi}{!}{
\fontsize{8}{10}\selectfont
\begin{tabular}[t]{lcccccccccccc}
\toprule
\multicolumn{1}{c}{ } & \multicolumn{4}{c}{Correct Group} & \multicolumn{4}{c}{Incorrect Group} & \multicolumn{4}{c}{Correct $-$ Incorrect} \\
\cmidrule(l{3pt}r{3pt}){2-5} \cmidrule(l{3pt}r{3pt}){6-9} \cmidrule(l{3pt}r{3pt}){10-13}
Country & $\beta_{1,1}$ & $\beta_{1,2}$ & $\beta_{1,3}$ & $\sum_{r=1}^{3}\beta_{1,r}$ & $\beta_{0,1}$ & $\beta_{0,2}$ & $\beta_{0,3}$ & $\sum_{r=1}^{3}\beta_{0,r}$ & $\Delta\beta_{1}$ & $\Delta\beta_{2}$ & $\Delta\beta_{3}$ & $\sum_{r=1}^{3}\Delta\beta_{r}$\\
\midrule
\cellcolor{gray!10}{AT} & \cellcolor{gray!10}{\(\textcolor{red}{-1.214}\)} & \cellcolor{gray!10}{\(\textcolor{blue}{0.594}\)} & \cellcolor{gray!10}{\(\textcolor{blue}{1.819}\)} & \cellcolor{gray!10}{\(\textcolor{blue}{1.200}\)} & \cellcolor{gray!10}{\(\textcolor{black}{-0.260}\)} & \cellcolor{gray!10}{\(\textcolor{black}{-0.095}\)} & \cellcolor{gray!10}{\(\textcolor{blue}{1.375}\)} & \cellcolor{gray!10}{\(\textcolor{blue}{1.020}\)} & \cellcolor{gray!10}{\(\textcolor{red}{-0.954}\)} & \cellcolor{gray!10}{\(\textcolor{blue}{0.689}\)} & \cellcolor{gray!10}{\(\textcolor{black}{0.444}\)} & \cellcolor{gray!10}{\(\textcolor{black}{0.180}\)}\\
BE & \(\textcolor{red}{-1.222}\) & \(\textcolor{blue}{0.805}\) & \(\textcolor{blue}{1.762}\) & \(\textcolor{blue}{1.345}\) & \(\textcolor{black}{-0.234}\) & \(\textcolor{black}{-0.121}\) & \(\textcolor{blue}{1.575}\) & \(\textcolor{blue}{1.220}\) & \(\textcolor{red}{-0.988}\) & \(\textcolor{blue}{0.926}\) & \(\textcolor{black}{0.187}\) & \(\textcolor{black}{0.125}\)\\
\cellcolor{gray!10}{DE} & \cellcolor{gray!10}{\(\textcolor{red}{-1.204}\)} & \cellcolor{gray!10}{\(\textcolor{blue}{0.575}\)} & \cellcolor{gray!10}{\(\textcolor{blue}{1.898}\)} & \cellcolor{gray!10}{\(\textcolor{blue}{1.269}\)} & \cellcolor{gray!10}{\(\textcolor{black}{-0.210}\)} & \cellcolor{gray!10}{\(\textcolor{black}{-0.080}\)} & \cellcolor{gray!10}{\(\textcolor{blue}{1.544}\)} & \cellcolor{gray!10}{\(\textcolor{blue}{1.254}\)} & \cellcolor{gray!10}{\(\textcolor{red}{-0.994}\)} & \cellcolor{gray!10}{\(\textcolor{blue}{0.655}\)} & \cellcolor{gray!10}{\(\textcolor{black}{0.354}\)} & \cellcolor{gray!10}{\(\textcolor{black}{0.015}\)}\\
DK & \(\textcolor{red}{-1.218}\) & \(\textcolor{blue}{0.611}\) & \(\textcolor{blue}{1.916}\) & \(\textcolor{blue}{1.309}\) & \(\textcolor{black}{-0.249}\) & \(\textcolor{black}{-0.051}\) & \(\textcolor{blue}{1.616}\) & \(\textcolor{blue}{1.317}\) & \(\textcolor{red}{-0.970}\) & \(\textcolor{blue}{0.662}\) & \(\textcolor{black}{0.300}\) & \(\textcolor{black}{-0.008}\)\\
\cellcolor{gray!10}{EE} & \cellcolor{gray!10}{\(\textcolor{red}{-1.217}\)} & \cellcolor{gray!10}{\(\textcolor{blue}{0.668}\)} & \cellcolor{gray!10}{\(\textcolor{blue}{1.661}\)} & \cellcolor{gray!10}{\(\textcolor{blue}{1.112}\)} & \cellcolor{gray!10}{\(\textcolor{black}{-0.256}\)} & \cellcolor{gray!10}{\(\textcolor{black}{0.034}\)} & \cellcolor{gray!10}{\(\textcolor{blue}{1.486}\)} & \cellcolor{gray!10}{\(\textcolor{blue}{1.264}\)} & \cellcolor{gray!10}{\(\textcolor{red}{-0.961}\)} & \cellcolor{gray!10}{\(\textcolor{blue}{0.634}\)} & \cellcolor{gray!10}{\(\textcolor{black}{0.175}\)} & \cellcolor{gray!10}{\(\textcolor{black}{-0.152}\)}\\
\addlinespace
FI & \(\textcolor{red}{-1.228}\) & \(\textcolor{blue}{0.462}\) & \(\textcolor{blue}{1.962}\) & \(\textcolor{blue}{1.196}\) & \(\textcolor{black}{-0.181}\) & \(\textcolor{black}{-0.155}\) & \(\textcolor{blue}{1.540}\) & \(\textcolor{blue}{1.204}\) & \(\textcolor{red}{-1.047}\) & \(\textcolor{blue}{0.617}\) & \(\textcolor{black}{0.422}\) & \(\textcolor{black}{-0.008}\)\\
\cellcolor{gray!10}{GB} & \cellcolor{gray!10}{\(\textcolor{red}{-1.203}\)} & \cellcolor{gray!10}{\(\textcolor{blue}{0.701}\)} & \cellcolor{gray!10}{\(\textcolor{blue}{1.835}\)} & \cellcolor{gray!10}{\(\textcolor{blue}{1.333}\)} & \cellcolor{gray!10}{\(\textcolor{black}{-0.206}\)} & \cellcolor{gray!10}{\(\textcolor{black}{-0.212}\)} & \cellcolor{gray!10}{\(\textcolor{blue}{1.512}\)} & \cellcolor{gray!10}{\(\textcolor{blue}{1.094}\)} & \cellcolor{gray!10}{\(\textcolor{red}{-0.997}\)} & \cellcolor{gray!10}{\(\textcolor{blue}{0.913}\)} & \cellcolor{gray!10}{\(\textcolor{black}{0.323}\)} & \cellcolor{gray!10}{\(\textcolor{black}{0.239}\)}\\
IE & \(\textcolor{red}{-1.212}\) & \(\textcolor{blue}{0.644}\) & \(\textcolor{blue}{1.920}\) & \(\textcolor{blue}{1.353}\) & \(\textcolor{black}{-0.201}\) & \(\textcolor{black}{-0.001}\) & \(\textcolor{blue}{1.425}\) & \(\textcolor{blue}{1.223}\) & \(\textcolor{red}{-1.010}\) & \(\textcolor{blue}{0.646}\) & \(\textcolor{black}{0.495}\) & \(\textcolor{black}{0.130}\)\\
\cellcolor{gray!10}{KR} & \cellcolor{gray!10}{\(\textcolor{red}{-1.234}\)} & \cellcolor{gray!10}{\(\textcolor{blue}{0.815}\)} & \cellcolor{gray!10}{\(\textcolor{blue}{1.771}\)} & \cellcolor{gray!10}{\(\textcolor{blue}{1.351}\)} & \cellcolor{gray!10}{\(\textcolor{black}{-0.236}\)} & \cellcolor{gray!10}{\(\textcolor{black}{0.018}\)} & \cellcolor{gray!10}{\(\textcolor{blue}{1.422}\)} & \cellcolor{gray!10}{\(\textcolor{blue}{1.204}\)} & \cellcolor{gray!10}{\(\textcolor{red}{-0.999}\)} & \cellcolor{gray!10}{\(\textcolor{blue}{0.797}\)} & \cellcolor{gray!10}{\(\textcolor{black}{0.348}\)} & \cellcolor{gray!10}{\(\textcolor{black}{0.147}\)}\\
NL & \(\textcolor{red}{-1.219}\) & \(\textcolor{blue}{0.727}\) & \(\textcolor{blue}{1.786}\) & \(\textcolor{blue}{1.294}\) & \(\textcolor{black}{-0.216}\) & \(\textcolor{black}{-0.095}\) & \(\textcolor{blue}{1.520}\) & \(\textcolor{blue}{1.208}\) & \(\textcolor{red}{-1.003}\) & \(\textcolor{blue}{0.823}\) & \(\textcolor{black}{0.266}\) & \(\textcolor{black}{0.086}\)\\
\addlinespace
\cellcolor{gray!10}{NO} & \cellcolor{gray!10}{\(\textcolor{red}{-1.204}\)} & \cellcolor{gray!10}{\(\textcolor{blue}{0.555}\)} & \cellcolor{gray!10}{\(\textcolor{blue}{1.929}\)} & \cellcolor{gray!10}{\(\textcolor{blue}{1.281}\)} & \cellcolor{gray!10}{\(\textcolor{black}{-0.190}\)} & \cellcolor{gray!10}{\(\textcolor{black}{-0.166}\)} & \cellcolor{gray!10}{\(\textcolor{blue}{1.741}\)} & \cellcolor{gray!10}{\(\textcolor{blue}{1.385}\)} & \cellcolor{gray!10}{\(\textcolor{red}{-1.014}\)} & \cellcolor{gray!10}{\(\textcolor{blue}{0.722}\)} & \cellcolor{gray!10}{\(\textcolor{black}{0.188}\)} & \cellcolor{gray!10}{\(\textcolor{black}{-0.104}\)}\\
PL & \(\textcolor{red}{-1.222}\) & \(\textcolor{blue}{0.653}\) & \(\textcolor{blue}{1.760}\) & \(\textcolor{blue}{1.191}\) & \(\textcolor{black}{-0.224}\) & \(\textcolor{black}{-0.073}\) & \(\textcolor{blue}{1.391}\) & \(\textcolor{blue}{1.094}\) & \(\textcolor{red}{-0.997}\) & \(\textcolor{blue}{0.726}\) & \(\textcolor{black}{0.369}\) & \(\textcolor{black}{0.097}\)\\
\cellcolor{gray!10}{SK} & \cellcolor{gray!10}{\(\textcolor{red}{-1.229}\)} & \cellcolor{gray!10}{\(\textcolor{blue}{0.823}\)} & \cellcolor{gray!10}{\(\textcolor{blue}{1.815}\)} & \cellcolor{gray!10}{\(\textcolor{blue}{1.409}\)} & \cellcolor{gray!10}{\(\textcolor{black}{-0.186}\)} & \cellcolor{gray!10}{\(\textcolor{black}{-0.130}\)} & \cellcolor{gray!10}{\(\textcolor{blue}{1.747}\)} & \cellcolor{gray!10}{\(\textcolor{blue}{1.432}\)} & \cellcolor{gray!10}{\(\textcolor{red}{-1.043}\)} & \cellcolor{gray!10}{\(\textcolor{blue}{0.953}\)} & \cellcolor{gray!10}{\(\textcolor{black}{0.068}\)} & \cellcolor{gray!10}{\(\textcolor{black}{-0.022}\)}\\
US & \(\textcolor{red}{-1.209}\) & \(\textcolor{blue}{0.604}\) & \(\textcolor{blue}{1.939}\) & \(\textcolor{blue}{1.334}\) & \(\textcolor{black}{-0.210}\) & \(\textcolor{black}{0.001}\) & \(\textcolor{blue}{1.620}\) & \(\textcolor{blue}{1.410}\) & \(\textcolor{red}{-0.999}\) & \(\textcolor{blue}{0.604}\) & \(\textcolor{black}{0.320}\) & \(\textcolor{black}{-0.075}\)\\
\bottomrule
\end{tabular}}
\end{table}

\paragraph{Baseline Transition Intensities} Table~\ref{tab:kappa-top5-cd} lists the five largest posterior mean baseline intensities $\kappa_{c,m,l}$ per response group. In both groups the transitions from \texttt{sch} and \texttt{ss\_sch} to \texttt{keypress} are largest with higher values in the correct group. The transition from \texttt{book\_m\_l} to \texttt{book\_m\_ok} is among the top five in both groups with similar magnitude. Complete results are given in the accompanying code repository. Because these parameters are baseline intensities rather than transition probabilities, the routing structure is examined through the derived networks in Section~\ref{sec:cdcross}.

\begin{table}[htb!]
\centering
\caption{Five largest posterior mean baseline hazards $\kappa_{c,m,l}$ by response group for CD Tally with 95\% HPD intervals. Log-scale summaries are computed from posterior draws of $\log\kappa$.}
\label{tab:kappa-top5-cd}
\begin{threeparttable}
\small
\begin{tabularx}{\textwidth}{
@{}
l
>{\raggedright\arraybackslash}X
>{\raggedright\arraybackslash}X
>{\raggedleft\arraybackslash}p{3.5cm}
>{\raggedleft\arraybackslash}p{3.75cm}
@{}}
\toprule
Group & From & To & $\kappa$ mean [95\% HPD] & $\log(\kappa)$ mean [95\% HPD] \\
\midrule
\multirow{5}{*}{Correct $(c=1)$} & \texttt{sch} & \texttt{keypress} & 123.7 [103.6, 145.4] & 4.81 [4.64, 4.98] \\
 & \texttt{ss\_sch} & \texttt{keypress} & 107.8 [96.8, 120.8] & 4.68 [4.57, 4.79] \\
 & \texttt{so\_ok} & \texttt{wb} & 72.3 [59.0, 86.6] & 4.28 [4.09, 4.47] \\
 & \texttt{book\_m\_l} & \texttt{book\_m\_ok} & 41.7 [7.2, 87.1] & 3.59 [2.49, 4.63] \\
 & \texttt{combobox} & \texttt{ss} & 38.4 [35.0, 41.5] & 3.65 [3.56, 3.73] \\
\addlinespace
\multirow{5}{*}{Incorrect $(c=0)$} & \texttt{ss\_sch} & \texttt{keypress} & 68.1 [55.0, 82.2] & 4.22 [4.01, 4.41] \\
 & \texttt{sch} & \texttt{keypress} & 57.8 [38.4, 77.8] & 4.04 [3.70, 4.39] \\
 & \texttt{book\_m\_l} & \texttt{book\_m\_ok} & 49.2 [15.5, 90.7] & 3.81 [2.97, 4.61] \\
 & \texttt{book\_a} & \texttt{book\_a\_ok} & 40.2 [19.0, 62.3] & 3.65 [3.11, 4.22] \\
 & \texttt{ss\_pst} & \texttt{ss\_so} & 33.0 [10.5, 58.3] & 3.42 [2.67, 4.19] \\
\bottomrule
\end{tabularx}
\begin{tablenotes}
\footnotesize
\item \textit{Note.}
Rows are ordered by posterior mean within each response group. $\kappa$ denotes the baseline transition hazard. Log-scale summaries were computed from posterior draws of $\log(\kappa)$, not by taking the logarithm of the posterior mean of $\kappa$.
\end{tablenotes}
\end{threeparttable}
\end{table}

\subsubsection{Transition-Network Comparisons Across Countries and Response Groups}\label{sec:cdcross}

We compare the 28 country-group networks of CD Tally at two levels. At the global level three distance-based analyses of the $G\times G$ posterior mean Wasserstein distance matrix $\bar D$ address whether the response groups differ in network structure and how much countries vary within each group. Non-metric multidimensional scaling (nMDS) \citep{kruskal1964multidimensional, kruskal1964nonmetric} visualizes $\bar D$ in two dimensions. Permutational multivariate analysis of variance (PERMANOVA) \citep{anderson2001new} tests whether the two response groups differ in location in the distance space and permutational analysis of multivariate dispersions (PERMDISP) \citep{anderson2006distance} tests whether they differ in within-group dispersion. At the local level, an additive and multiplicative effects (AME) network model \citep{hoff2009multiplicative, hoff2021additive} of the difference between two selected networks identifies the actions and pathways that account for their dissimilarity.

\paragraph{Embedding and Distances} The shared DiGAE encoder was trained on the $G=28$ posterior mean networks with $d=16$, $d_1=32$, $\gamma^{+}=0.1$, and $\gamma^{-}=0.6$. The selection procedure for these hyperparameters is described in the Section~S3.1 of the Supplementary Material. Applied to the 280,000 posterior network draws the fixed encoder reduced the row-averaged KL divergence relative to a support-restricted uniform baseline by 83.7\% to 92.7\% across combinations with a mean of 88.9\% and preserved the ranking of dominant transitions with a mean normalized discounted cumulative gain at rank 5 (NDCG@5) of 0.971 \citep{jarvelin2002cumulated}. Full diagnostics are given in the Section~S3.3 of the Supplementary Material.

\begin{figure}[htb!]
    \centering
    \includegraphics[width=0.75\linewidth]{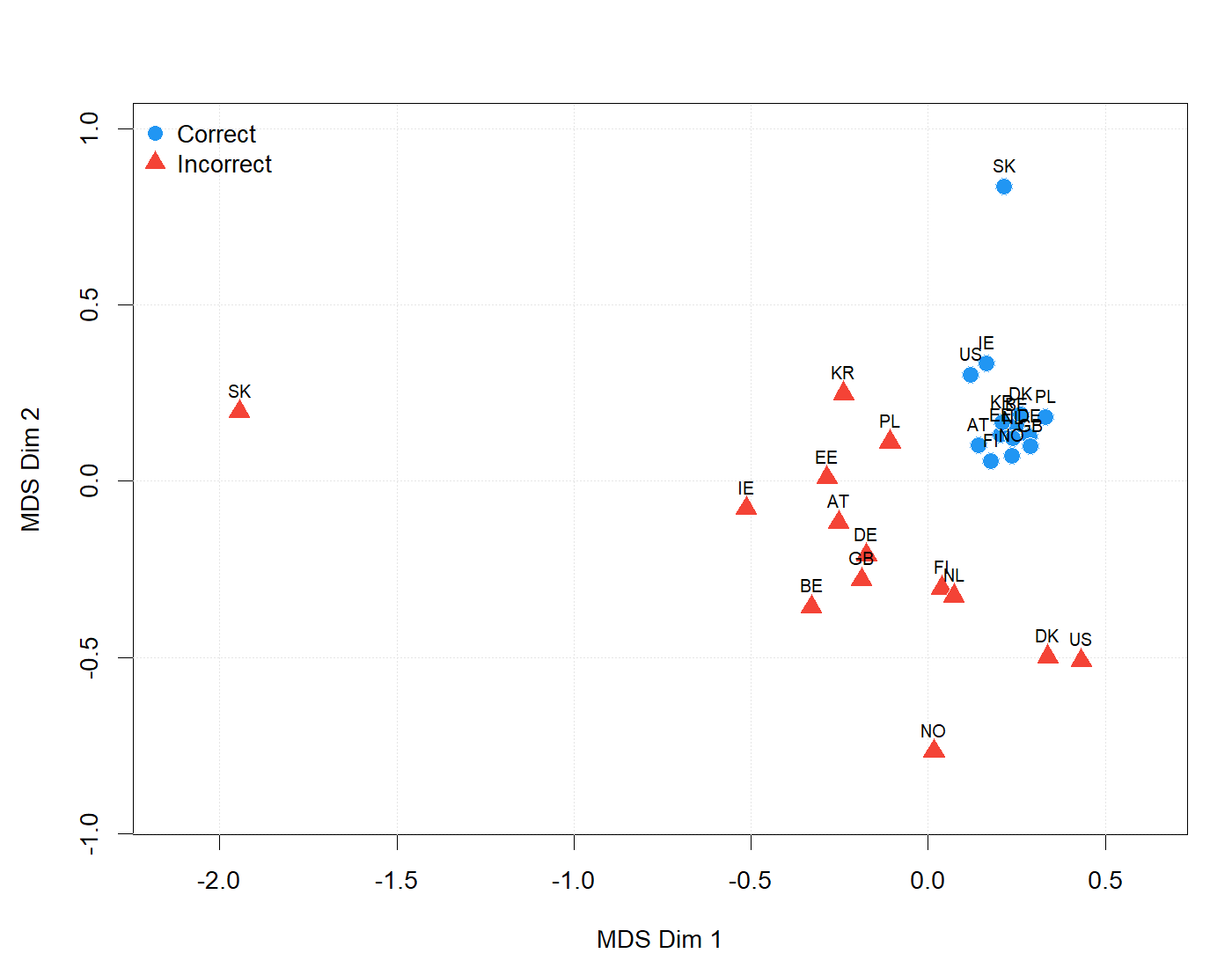}
    \caption{nMDS representation of the 28 CD Tally networks based on $\bar D$ (Stress-1 = 0.116). Circles are correct networks and triangles are incorrect networks.}
    \label{fig:cd_mds}
\end{figure}

\begin{figure}[htb!]
    \centering
    \begin{subfigure}[t]{0.43\linewidth}
        \centering
        \includegraphics[width=\linewidth]{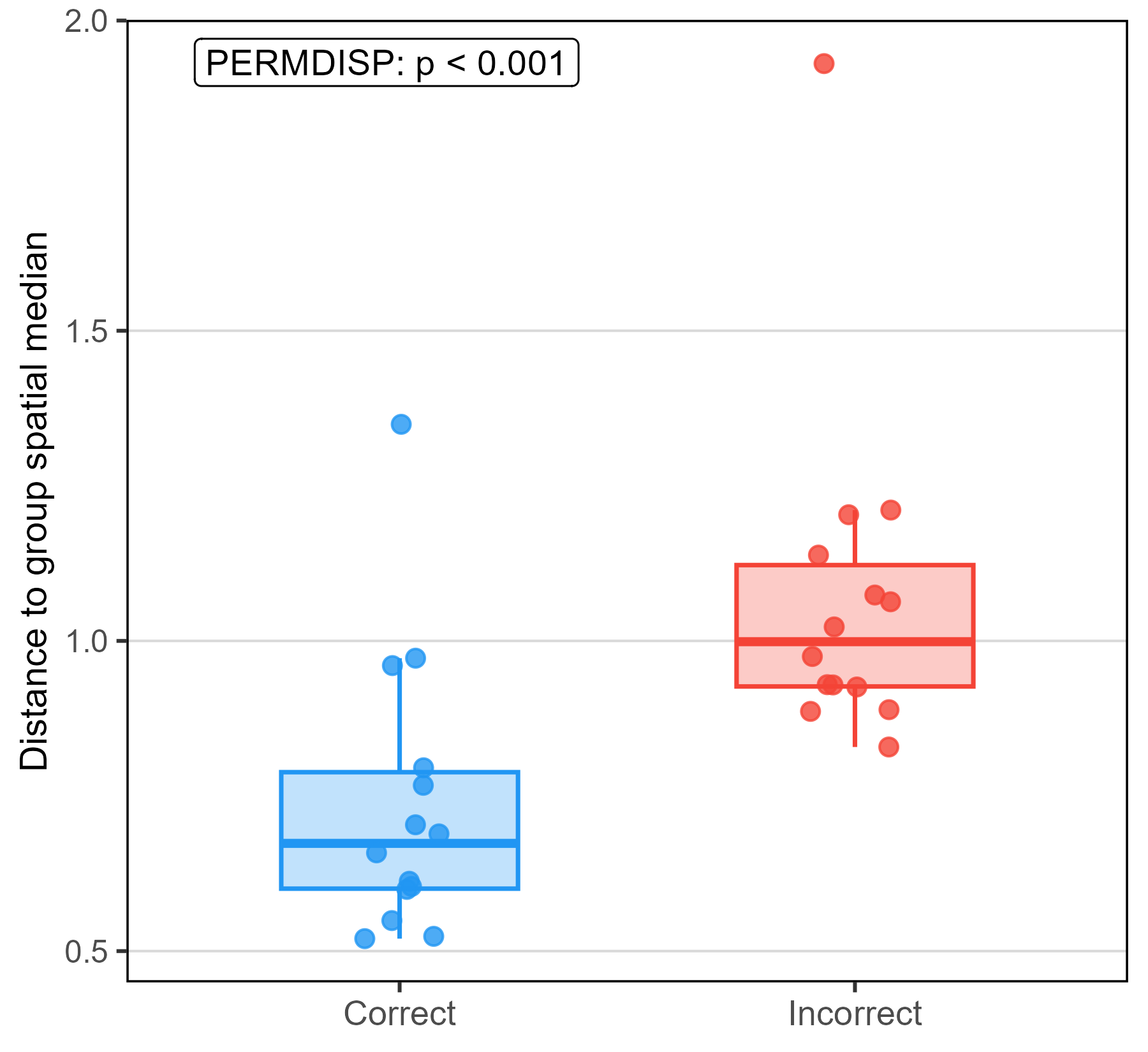}
        \caption{}
        \label{fig:cd_dispersion_group}
    \end{subfigure}
    \hfill
    \begin{subfigure}[t]{0.4\linewidth}
        \centering
        \includegraphics[width=\linewidth]{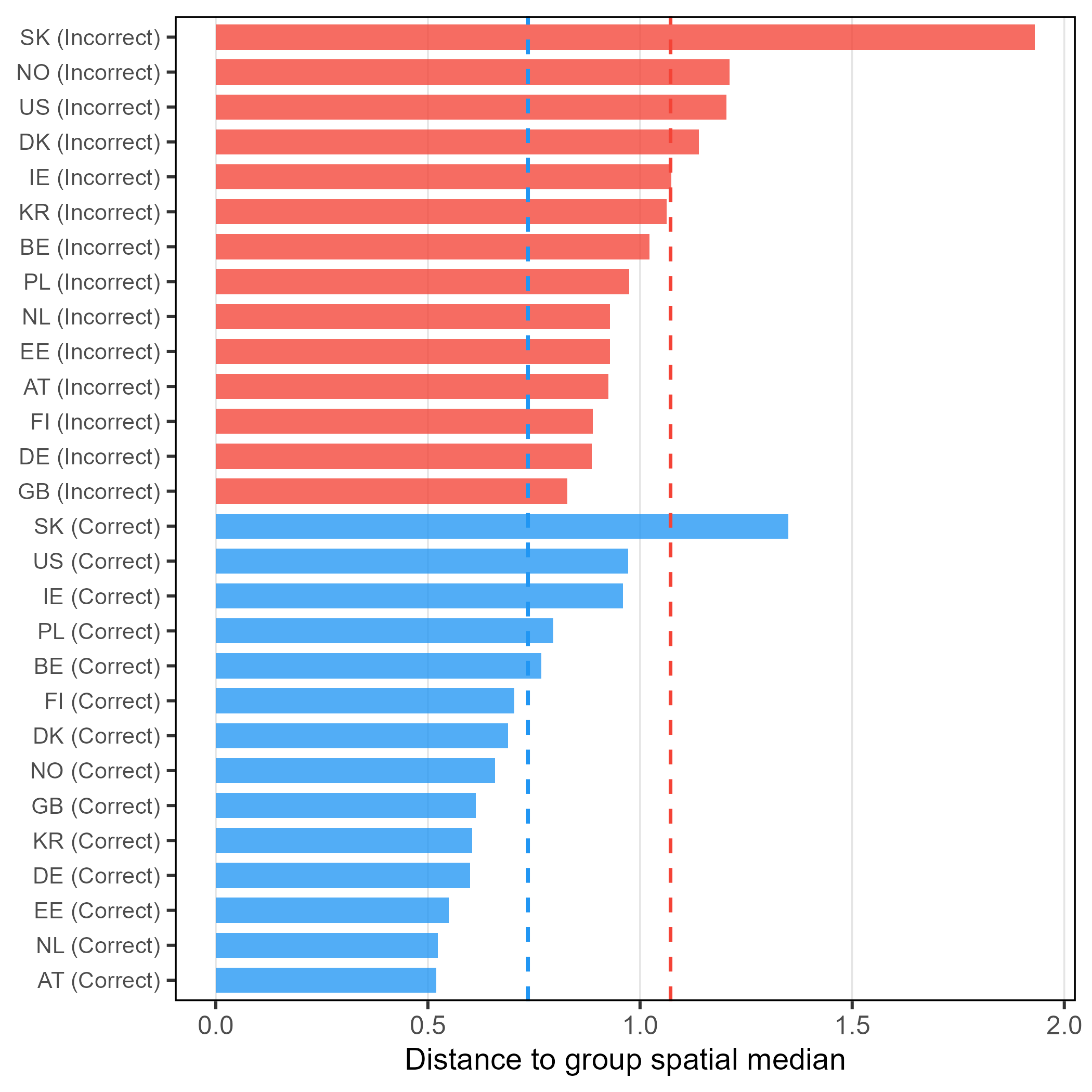}
        \caption{}
        \label{fig:cd_dispersion_graph}
    \end{subfigure}
    \caption{Dispersion of the CD Tally networks by response group. (a) Distances of the country networks to their group spatial median. (b) The same distances by network. Dashed lines mark group means.}
    \label{fig:cd_dispersion}
\end{figure}

\paragraph{Global Differences} Figure~\ref{fig:cd_mds} shows the two-dimensional nMDS solution of $\bar D$ (Stress-1 = 0.116). Correct and incorrect networks occupy partly distinct regions and the correct networks form a more compact cluster. PERMANOVA with permutations restricted within country \citep{anderson2003permutation} indicated a difference between the two response groups ($F(1,26)=6.588$, $R^2=0.202$, $p<0.001$). PERMDISP indicated a difference in within-group dispersion ($F(1,26)=12.422$, $p<0.001$) with a larger mean distance to the group spatial median for incorrect networks (1.072 versus 0.736) as shown in Figure~\ref{fig:cd_dispersion}. The PERMANOVA result therefore reflects both a shift in location and greater cross-country heterogeneity among incorrect networks \citep{anderson2013permanova}. Correct responses were associated with a routing structure that is shared across countries whereas incorrect responses were associated with more country-specific structures.

\paragraph{Local Structure} To relate the global differences to specific transitions we applied an additive and multiplicative effects (AME) network model \citep{hoff2009multiplicative, hoff2021additive} to two contrasts: the correct and incorrect networks of Austria (AT) and the incorrect networks of Slovakia (SK) and United Kingdom (GB) where GB is central and SK peripheral within the incorrect group. For each contrast the differential network $\Delta$ is the difference between the posterior mean transition matrices with positive entries indicating transitions that are relatively stronger in the first network. Source actions observed in both networks are retained and target actions are the union of observed targets. The decomposition $\Delta_{ml}=\mu+a_m+b_l+u_m^\top v_l+\epsilon_{ml}$ separates a sender effect $a_m$, a receiver effect $b_l$, and a source-target component $u_m^\top v_l$. Source actions observed in only one network are summarized separately. Details are given in the Section~S5 of the Supplementary Material.

\begin{figure}[htb!]
    \centering
    \begin{subfigure}[t]{0.48\linewidth}
        \centering
        \includegraphics[width=\linewidth]
        {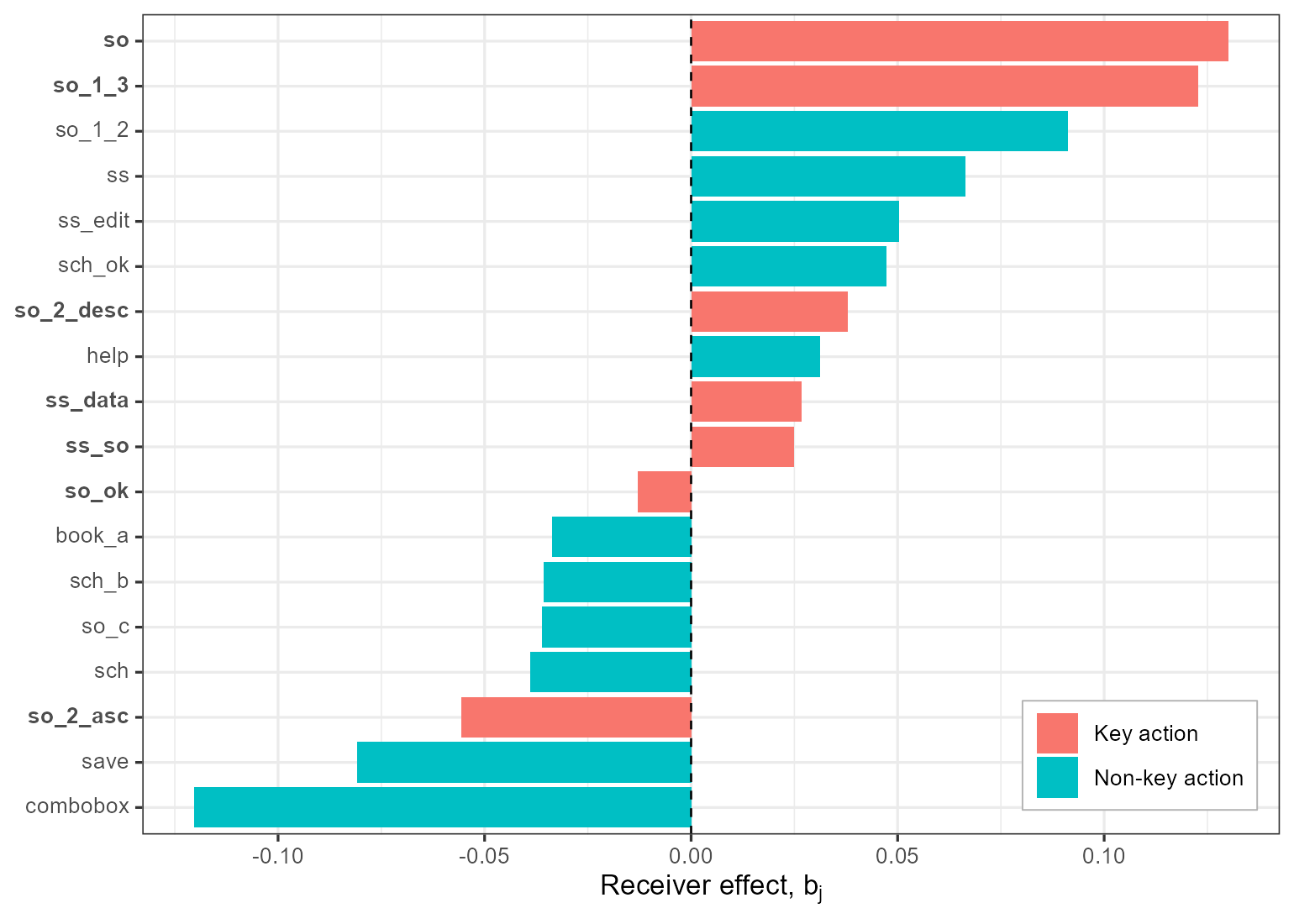}
        \caption{}
        \label{fig:ame_at_receiver}
    \end{subfigure}
    \hfill
    \begin{subfigure}[t]{0.48\linewidth}
        \centering
        \includegraphics[width=\linewidth]
        {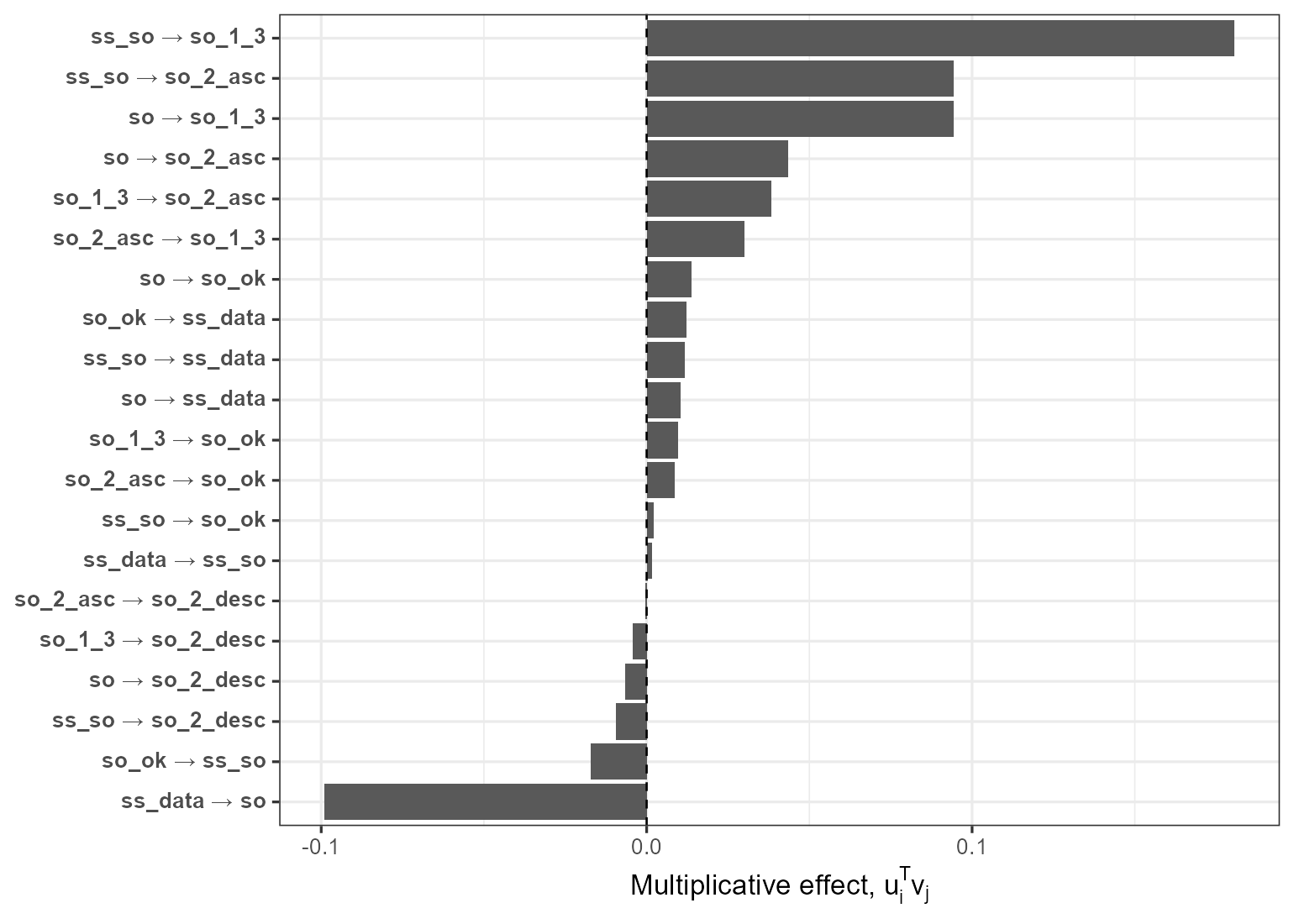}
        \caption{}
        \label{fig:ame_at_mult}
    \end{subfigure}

    \vspace{0.5em}

    \begin{subfigure}[t]{0.48\linewidth}
        \centering
        \includegraphics[width=\linewidth]
        {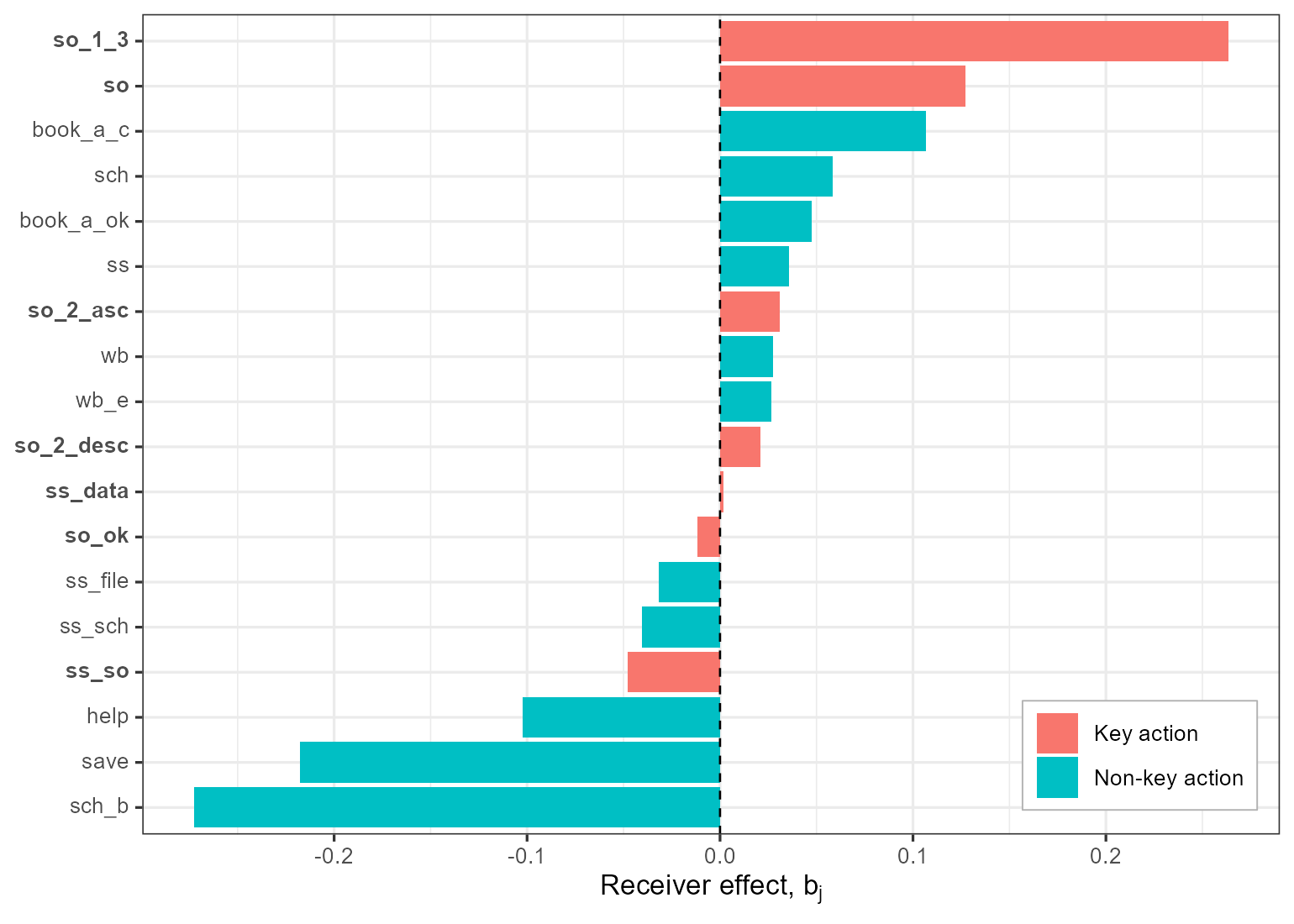}
        \caption{}
        \label{fig:ame_skgb_receiver}
    \end{subfigure}
    \hfill
    \begin{subfigure}[t]{0.48\linewidth}
        \centering
        \includegraphics[width=\linewidth]
        {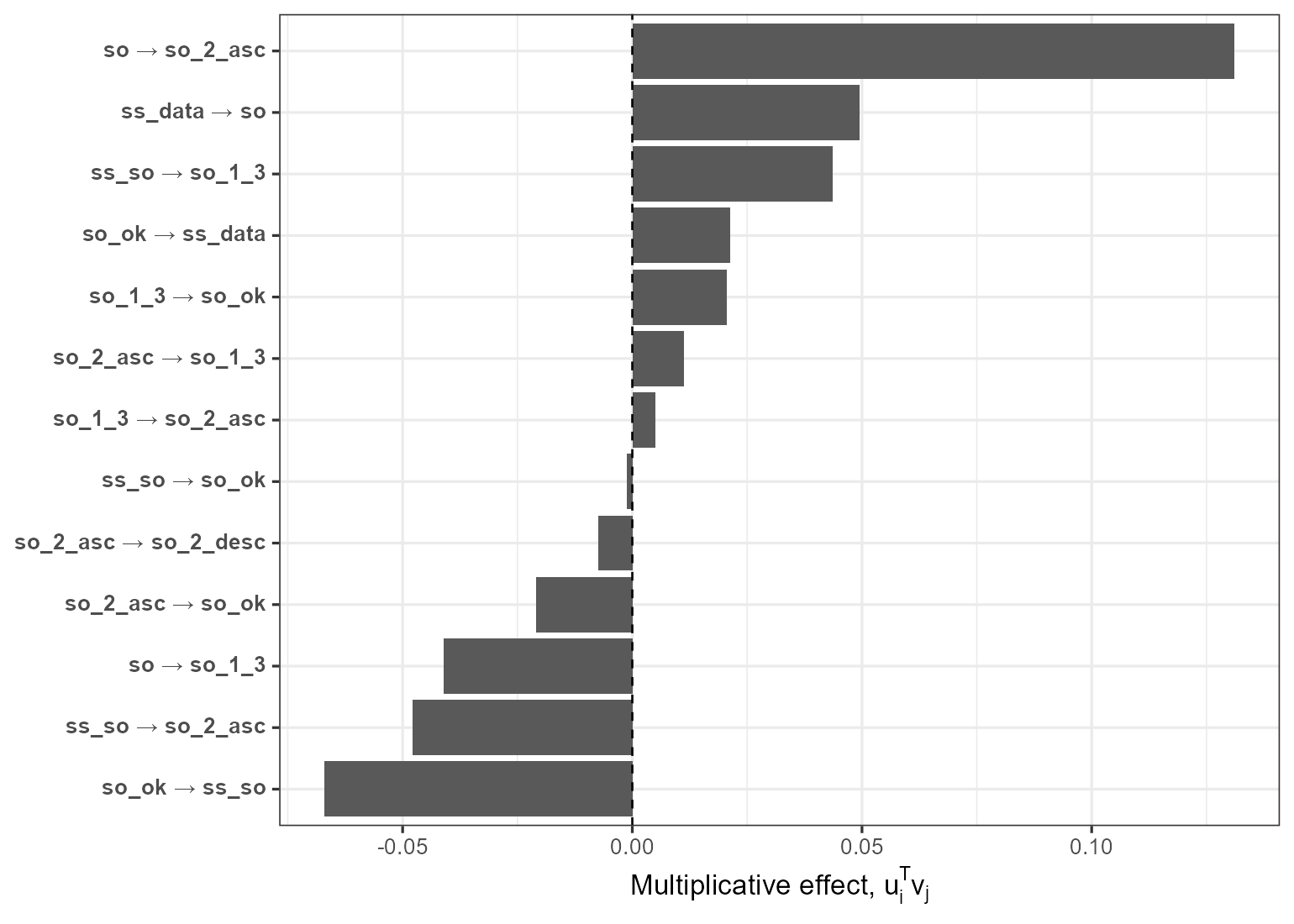}
        \caption{}
        \label{fig:ame_skgb_mult}
    \end{subfigure}

    \caption{AME decomposition of two CD Tally contrasts. Panels (a) and (b): correct versus incorrect networks of AT. Panels (c) and (d): incorrect networks of SK versus GB. Panels (a) and (c) show receiver effects $b_l$ for the 15 actions with the largest absolute effects together with all key actions (highlighted). Panels (b) and (d) show multiplicative effects $u_m^\top v_l$ for observed pathways between key actions. Positive values are oriented toward the first network of each contrast.}
    \label{fig:ame}
\end{figure}

\begin{itemize}
\item {\bf Austria: Correct versus Incorrect} Panels (a) and (b) of Figure~\ref{fig:ame} show the receiver effects and the multiplicative effects among key actions. Among key actions \texttt{so} ($b_l=0.130$) and \texttt{so\_1\_3} ($b_l=0.123$) have the largest positive receiver effects so that incoming probability is allocated more toward these actions in the correct network whereas \texttt{so\_2\_asc} has a negative effect ($b_l=-0.056$). The multiplicative component identifies pathway differences beyond the receiver effects. The pathway \texttt{ss\_so} $\rightarrow$ \texttt{so\_1\_3} (entering the sort dialog and selecting genre) is correct-oriented ($u_m^\top v_l=0.181$, $\Delta_{ml}=0.455$). Following confirmation, \texttt{so\_ok} $\rightarrow$ \texttt{wb} (returning to the website where the answer is entered) is correct-oriented ($u_m^\top v_l=0.089$, $\Delta_{ml}=0.145$) whereas \texttt{so\_ok} $\rightarrow$ \texttt{ss\_so} (returning to the sort dialog) is incorrect-oriented ($u_m^\top v_l=-0.017$, $\Delta_{ml}=-0.073$). The two groups thus differ in what follows the same sorting confirmation step. Support also differs. The key action \texttt{so\_2\_desc} appears as a source only in the correct network with most of its probability on \texttt{so\_ok} (0.782) and \texttt{ss\_cut} appears as a source only in the incorrect network with the largest probability on \texttt{ss\_edit} (0.599).

\item {\bf Slovakia versus the United Kingdom within the Incorrect Group} Panels (c) and (d) of Figure~\ref{fig:ame} show the SK versus GB contrast. Among key actions \texttt{so\_1\_3} has the largest positive receiver effect ($b_l=0.263$) toward SK and \texttt{ss\_so} a negative effect ($b_l=-0.048$) toward GB. The pathway \texttt{ss\_so} $\rightarrow$ \texttt{so\_1\_3} is SK-oriented ($u_m^\top v_l=0.044$, $\Delta_{ml}=0.304$) and \texttt{so\_ok} $\rightarrow$ \texttt{ss\_so} (returning to the sort dialog after confirmation) is GB-oriented ($u_m^\top v_l=-0.067$, $\Delta_{ml}=-0.136$). The genre-sorting sequence thus contributes to both contrasts although its meaning depends on the surrounding network. The contrast also shows an asymmetry in support. Eleven source actions appear only in GB and none only in SK so that the separation of SK reflects both different routing among shared sources and a narrower set of source actions.
\end{itemize}

\subsection{Lamp Return Item}\label{sec:lamp}

\subsubsection{Parameter Estimation}\label{sec:lamppara}

\begin{figure}[htb!]
    \centering
    \includegraphics[width=0.9\linewidth]{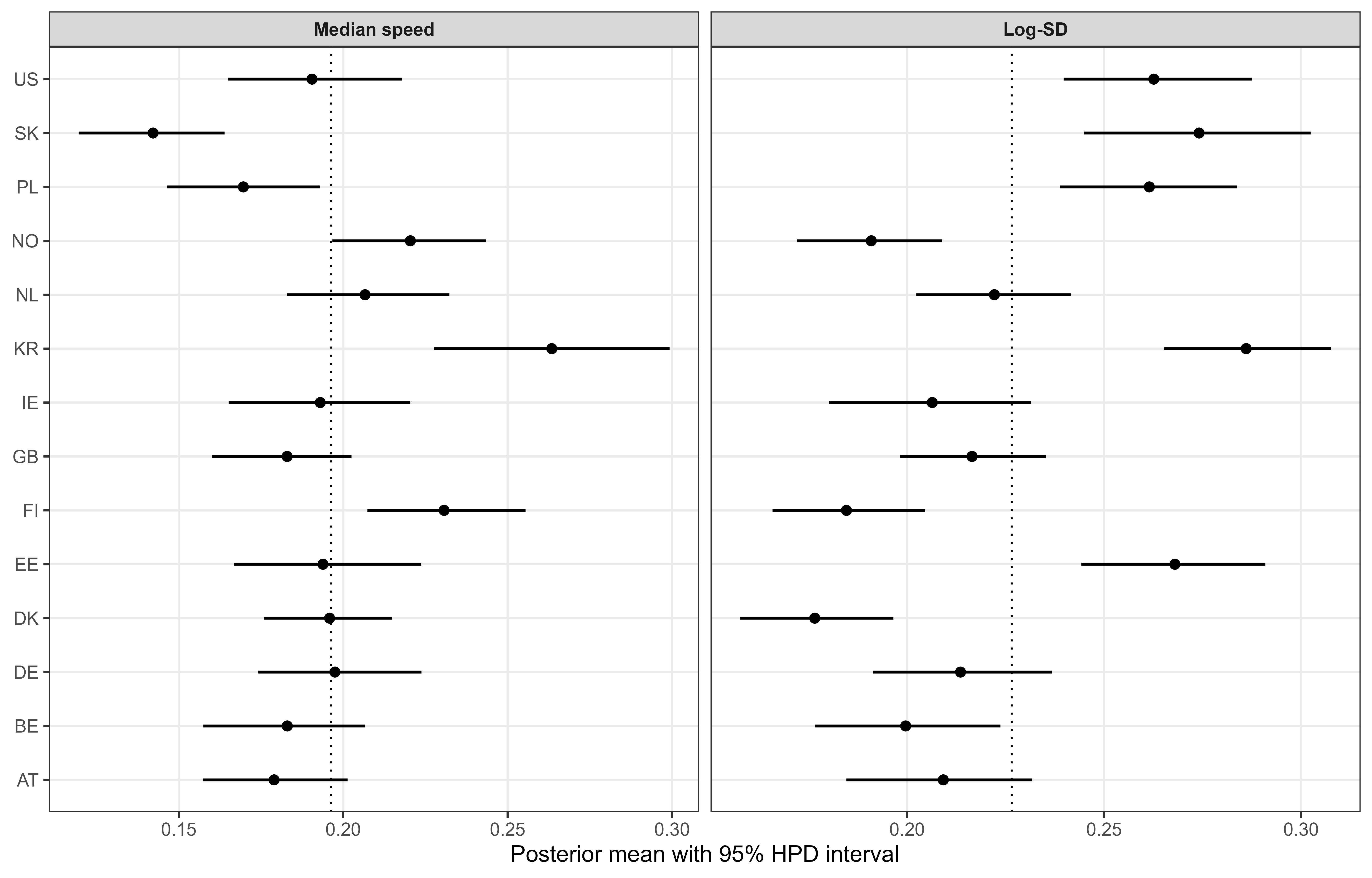}
    \caption{Posterior means and 95\% HPD intervals of the country-level median speed $\exp(\mu_{\tau,q})$ (left) and the within-country standard deviation of log speed $\sigma_{\tau,q}$ (right) for Lamp Return. Dotted lines mark the means across countries.}
    \label{fig:lamp_tau}
\end{figure}

\paragraph{Country-level Speed} Figure~\ref{fig:lamp_tau} shows the country-level speed parameters. South Korea has the highest posterior mean of $\exp(\mu_{\tau,q})$ and Slovakia the lowest and their 95\% HPD intervals do not overlap. The within-country standard deviation of log speed is largest for South Korea and Slovakia and smallest for Denmark and Finland. Slovakia is slowest in both items whereas the fastest country differs (Finland for CD Tally and South Korea for Lamp Return).

\begin{figure}[htb!]
    \centering
    \includegraphics[width=1\linewidth]{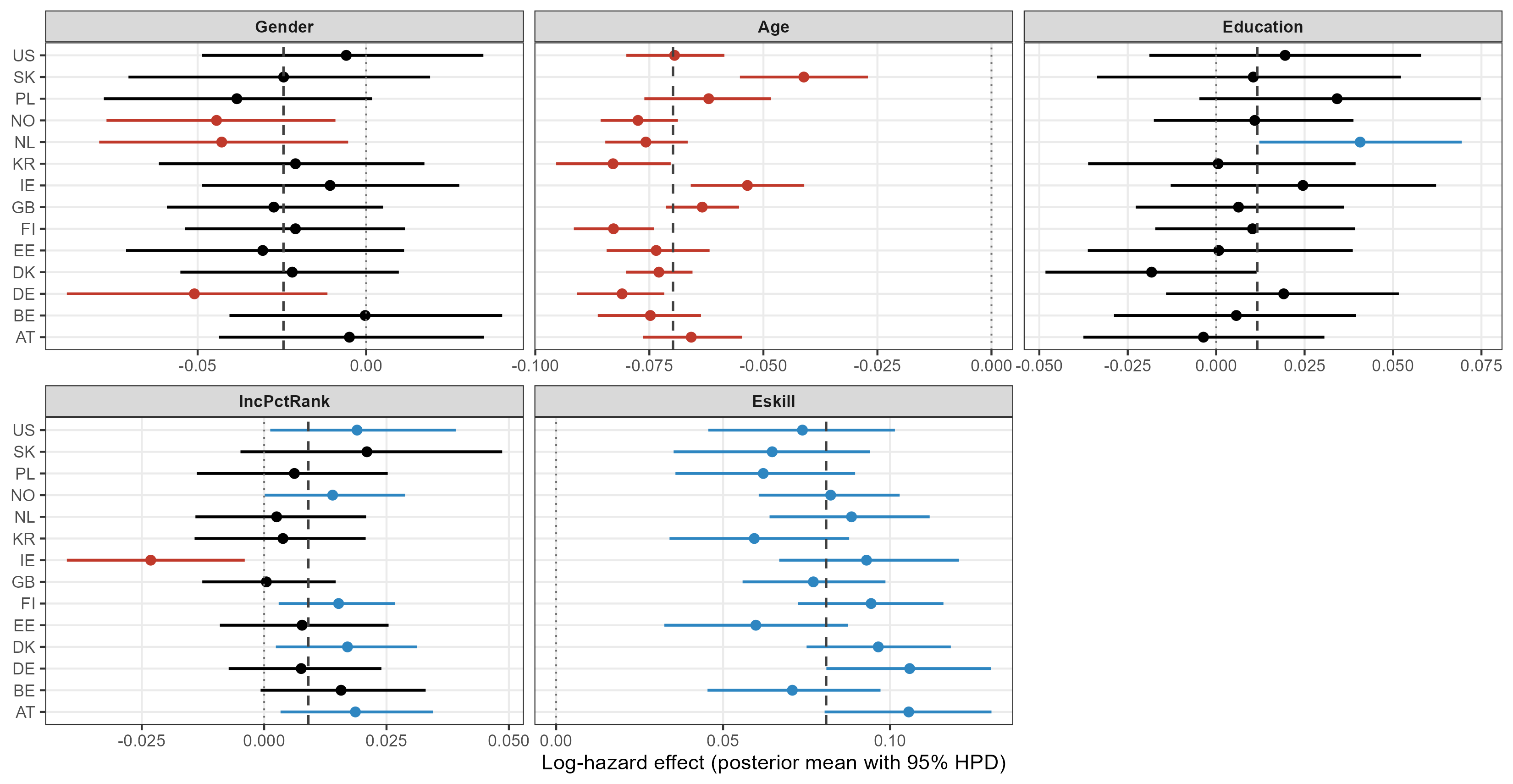}
    \caption{Country-specific covariate effects $\alpha_{p,q}$ on the log-hazard for Lamp Return. Points are posterior means and bars are 95\% HPD intervals. Blue and red mark intervals above and below zero and black marks intervals that include zero. Dashed lines mark the pooled effects $\mu_{\alpha,p}$ and dotted lines mark zero.}
    \label{fig:lamp_alpha}
\end{figure}

\paragraph{Covariate Effects} Figure~\ref{fig:lamp_alpha} shows the covariate effects. Age is credibly negative and Eskill credibly positive in all 14 countries as in CD Tally. Gender is negative in all countries and credibly negative only in DE, NL, and NO. Education is mostly positive and credibly positive only in NL. Income is credibly positive in AT, DK, FI, NO, and US and credibly negative in IE. Age and Eskill are thus consistent across items whereas gender, education, and income vary by item and country.

\paragraph{Key-action Effects} Table~\ref{tab:key_action_effects_lamp} reports the key-action effects. In the correct group $\beta_{1,1,q}$ is credibly negative and $\beta_{1,2,q}$, $\beta_{1,3,q}$, and the total effect are credibly positive in all 14 countries. Unlike CD Tally the incorrect group shows the same pattern: $\beta_{0,1,q}$ is credibly negative and $\beta_{0,2,q}$ and $\beta_{0,3,q}$ are credibly positive in every country and the total effect is credibly positive in 13 countries with the interval including zero only in South Korea. The contrasts also differ from CD Tally. $\Delta\beta_{1,q}$ is credibly negative and $\Delta\beta_{3,q}$ credibly positive in all countries whereas the intervals of $\Delta\beta_{2,q}$ and of the total contrast include zero throughout. CD Tally thus separates the groups through both main effects whereas Lamp Return separates them through a stronger negative source effect and a stronger key-to-key interaction in the correct group. In neither item does the total key-to-key effect differ credibly between groups.

\begin{table}[htb!]
\centering
\caption{Posterior means of key-action effects by country for Lamp Return. The total effect $\sum_{r=1}^{3}\beta_{c,r,q}$ is the log-hazard effect of a key-to-key transition. The contrast is $\Delta\beta_{r,q}=\beta_{1,r,q}-\beta_{0,r,q}$. Blue and red mark 95\% HPD intervals above and below zero and black marks intervals that include zero.}
\label{tab:key_action_effects_lamp}
\centering
\resizebox{\ifdim\width>\linewidth\linewidth\else\width\fi}{!}{
\fontsize{8}{10}\selectfont
\begin{tabular}[t]{lcccccccccccc}
\toprule
\multicolumn{1}{c}{ } & \multicolumn{4}{c}{Correct Group} & \multicolumn{4}{c}{Incorrect Group} & \multicolumn{4}{c}{Correct $-$ Incorrect} \\
\cmidrule(l{3pt}r{3pt}){2-5} \cmidrule(l{3pt}r{3pt}){6-9} \cmidrule(l{3pt}r{3pt}){10-13}
Country & $\beta_{1,1}$ & $\beta_{1,2}$ & $\beta_{1,3}$ & $\sum_{r=1}^{3}\beta_{1,r}$ & $\beta_{0,1}$ & $\beta_{0,2}$ & $\beta_{0,3}$ & $\sum_{r=1}^{3}\beta_{0,r}$ & $\Delta\beta_{1}$ & $\Delta\beta_{2}$ & $\Delta\beta_{3}$ & $\sum_{r=1}^{3}\Delta\beta_{r}$\\
\midrule
\cellcolor{gray!10}{AT} & \cellcolor{gray!10}{\(\textcolor{red}{-3.120}\)} & \cellcolor{gray!10}{\(\textcolor{blue}{1.352}\)} & \cellcolor{gray!10}{\(\textcolor{blue}{2.320}\)} & \cellcolor{gray!10}{\(\textcolor{blue}{0.552}\)} & \cellcolor{gray!10}{\(\textcolor{red}{-2.396}\)} & \cellcolor{gray!10}{\(\textcolor{blue}{1.297}\)} & \cellcolor{gray!10}{\(\textcolor{blue}{1.411}\)} & \cellcolor{gray!10}{\(\textcolor{blue}{0.312}\)} & \cellcolor{gray!10}{\(\textcolor{red}{-0.724}\)} & \cellcolor{gray!10}{\(\textcolor{black}{0.055}\)} & \cellcolor{gray!10}{\(\textcolor{blue}{0.909}\)} & \cellcolor{gray!10}{\(\textcolor{black}{0.240}\)}\\
BE & \(\textcolor{red}{-3.183}\) & \(\textcolor{blue}{1.398}\) & \(\textcolor{blue}{2.384}\) & \(\textcolor{blue}{0.599}\) & \(\textcolor{red}{-2.405}\) & \(\textcolor{blue}{1.386}\) & \(\textcolor{blue}{1.514}\) & \(\textcolor{blue}{0.495}\) & \(\textcolor{red}{-0.778}\) & \(\textcolor{black}{0.011}\) & \(\textcolor{blue}{0.871}\) & \(\textcolor{black}{0.104}\)\\
\cellcolor{gray!10}{DE} & \cellcolor{gray!10}{\(\textcolor{red}{-3.113}\)} & \cellcolor{gray!10}{\(\textcolor{blue}{1.337}\)} & \cellcolor{gray!10}{\(\textcolor{blue}{2.344}\)} & \cellcolor{gray!10}{\(\textcolor{blue}{0.568}\)} & \cellcolor{gray!10}{\(\textcolor{red}{-2.381}\)} & \cellcolor{gray!10}{\(\textcolor{blue}{1.304}\)} & \cellcolor{gray!10}{\(\textcolor{blue}{1.442}\)} & \cellcolor{gray!10}{\(\textcolor{blue}{0.365}\)} & \cellcolor{gray!10}{\(\textcolor{red}{-0.732}\)} & \cellcolor{gray!10}{\(\textcolor{black}{0.033}\)} & \cellcolor{gray!10}{\(\textcolor{blue}{0.902}\)} & \cellcolor{gray!10}{\(\textcolor{black}{0.203}\)}\\
DK & \(\textcolor{red}{-3.128}\) & \(\textcolor{blue}{1.290}\) & \(\textcolor{blue}{2.453}\) & \(\textcolor{blue}{0.616}\) & \(\textcolor{red}{-2.364}\) & \(\textcolor{blue}{1.336}\) & \(\textcolor{blue}{1.519}\) & \(\textcolor{blue}{0.491}\) & \(\textcolor{red}{-0.764}\) & \(\textcolor{black}{-0.045}\) & \(\textcolor{blue}{0.934}\) & \(\textcolor{black}{0.125}\)\\
\cellcolor{gray!10}{EE} & \cellcolor{gray!10}{\(\textcolor{red}{-3.029}\)} & \cellcolor{gray!10}{\(\textcolor{blue}{1.317}\)} & \cellcolor{gray!10}{\(\textcolor{blue}{2.183}\)} & \cellcolor{gray!10}{\(\textcolor{blue}{0.471}\)} & \cellcolor{gray!10}{\(\textcolor{red}{-2.363}\)} & \cellcolor{gray!10}{\(\textcolor{blue}{1.226}\)} & \cellcolor{gray!10}{\(\textcolor{blue}{1.421}\)} & \cellcolor{gray!10}{\(\textcolor{blue}{0.283}\)} & \cellcolor{gray!10}{\(\textcolor{red}{-0.666}\)} & \cellcolor{gray!10}{\(\textcolor{black}{0.091}\)} & \cellcolor{gray!10}{\(\textcolor{blue}{0.763}\)} & \cellcolor{gray!10}{\(\textcolor{black}{0.188}\)}\\
\addlinespace
FI & \(\textcolor{red}{-3.090}\) & \(\textcolor{blue}{1.207}\) & \(\textcolor{blue}{2.282}\) & \(\textcolor{blue}{0.399}\) & \(\textcolor{red}{-2.351}\) & \(\textcolor{blue}{1.161}\) & \(\textcolor{blue}{1.530}\) & \(\textcolor{blue}{0.340}\) & \(\textcolor{red}{-0.739}\) & \(\textcolor{black}{0.046}\) & \(\textcolor{blue}{0.752}\) & \(\textcolor{black}{0.059}\)\\
\cellcolor{gray!10}{GB} & \cellcolor{gray!10}{\(\textcolor{red}{-3.109}\)} & \cellcolor{gray!10}{\(\textcolor{blue}{1.355}\)} & \cellcolor{gray!10}{\(\textcolor{blue}{2.414}\)} & \cellcolor{gray!10}{\(\textcolor{blue}{0.660}\)} & \cellcolor{gray!10}{\(\textcolor{red}{-2.369}\)} & \cellcolor{gray!10}{\(\textcolor{blue}{1.399}\)} & \cellcolor{gray!10}{\(\textcolor{blue}{1.397}\)} & \cellcolor{gray!10}{\(\textcolor{blue}{0.427}\)} & \cellcolor{gray!10}{\(\textcolor{red}{-0.740}\)} & \cellcolor{gray!10}{\(\textcolor{black}{-0.044}\)} & \cellcolor{gray!10}{\(\textcolor{blue}{1.017}\)} & \cellcolor{gray!10}{\(\textcolor{black}{0.233}\)}\\
IE & \(\textcolor{red}{-3.149}\) & \(\textcolor{blue}{1.372}\) & \(\textcolor{blue}{2.390}\) & \(\textcolor{blue}{0.613}\) & \(\textcolor{red}{-2.341}\) & \(\textcolor{blue}{1.227}\) & \(\textcolor{blue}{1.508}\) & \(\textcolor{blue}{0.394}\) & \(\textcolor{red}{-0.808}\) & \(\textcolor{black}{0.145}\) & \(\textcolor{blue}{0.882}\) & \(\textcolor{black}{0.219}\)\\
\cellcolor{gray!10}{KR} & \cellcolor{gray!10}{\(\textcolor{red}{-2.866}\)} & \cellcolor{gray!10}{\(\textcolor{blue}{1.132}\)} & \cellcolor{gray!10}{\(\textcolor{blue}{2.156}\)} & \cellcolor{gray!10}{\(\textcolor{blue}{0.422}\)} & \cellcolor{gray!10}{\(\textcolor{red}{-2.284}\)} & \cellcolor{gray!10}{\(\textcolor{blue}{1.124}\)} & \cellcolor{gray!10}{\(\textcolor{blue}{1.327}\)} & \cellcolor{gray!10}{\(\textcolor{black}{0.167}\)} & \cellcolor{gray!10}{\(\textcolor{red}{-0.582}\)} & \cellcolor{gray!10}{\(\textcolor{black}{0.008}\)} & \cellcolor{gray!10}{\(\textcolor{blue}{0.829}\)} & \cellcolor{gray!10}{\(\textcolor{black}{0.255}\)}\\
NL & \(\textcolor{red}{-3.133}\) & \(\textcolor{blue}{1.352}\) & \(\textcolor{blue}{2.368}\) & \(\textcolor{blue}{0.587}\) & \(\textcolor{red}{-2.404}\) & \(\textcolor{blue}{1.368}\) & \(\textcolor{blue}{1.526}\) & \(\textcolor{blue}{0.491}\) & \(\textcolor{red}{-0.729}\) & \(\textcolor{black}{-0.017}\) & \(\textcolor{blue}{0.842}\) & \(\textcolor{black}{0.096}\)\\
\addlinespace
\cellcolor{gray!10}{NO} & \cellcolor{gray!10}{\(\textcolor{red}{-3.116}\)} & \cellcolor{gray!10}{\(\textcolor{blue}{1.273}\)} & \cellcolor{gray!10}{\(\textcolor{blue}{2.310}\)} & \cellcolor{gray!10}{\(\textcolor{blue}{0.466}\)} & \cellcolor{gray!10}{\(\textcolor{red}{-2.387}\)} & \cellcolor{gray!10}{\(\textcolor{blue}{1.337}\)} & \cellcolor{gray!10}{\(\textcolor{blue}{1.465}\)} & \cellcolor{gray!10}{\(\textcolor{blue}{0.415}\)} & \cellcolor{gray!10}{\(\textcolor{red}{-0.730}\)} & \cellcolor{gray!10}{\(\textcolor{black}{-0.064}\)} & \cellcolor{gray!10}{\(\textcolor{blue}{0.845}\)} & \cellcolor{gray!10}{\(\textcolor{black}{0.051}\)}\\
PL & \(\textcolor{red}{-3.025}\) & \(\textcolor{blue}{1.334}\) & \(\textcolor{blue}{2.296}\) & \(\textcolor{blue}{0.604}\) & \(\textcolor{red}{-2.360}\) & \(\textcolor{blue}{1.285}\) & \(\textcolor{blue}{1.513}\) & \(\textcolor{blue}{0.438}\) & \(\textcolor{red}{-0.665}\) & \(\textcolor{black}{0.048}\) & \(\textcolor{blue}{0.782}\) & \(\textcolor{black}{0.166}\)\\
\cellcolor{gray!10}{SK} & \cellcolor{gray!10}{\(\textcolor{red}{-3.114}\)} & \cellcolor{gray!10}{\(\textcolor{blue}{1.398}\)} & \cellcolor{gray!10}{\(\textcolor{blue}{2.338}\)} & \cellcolor{gray!10}{\(\textcolor{blue}{0.622}\)} & \cellcolor{gray!10}{\(\textcolor{red}{-2.322}\)} & \cellcolor{gray!10}{\(\textcolor{blue}{1.287}\)} & \cellcolor{gray!10}{\(\textcolor{blue}{1.505}\)} & \cellcolor{gray!10}{\(\textcolor{blue}{0.471}\)} & \cellcolor{gray!10}{\(\textcolor{red}{-0.792}\)} & \cellcolor{gray!10}{\(\textcolor{black}{0.111}\)} & \cellcolor{gray!10}{\(\textcolor{blue}{0.832}\)} & \cellcolor{gray!10}{\(\textcolor{black}{0.151}\)}\\
US & \(\textcolor{red}{-3.146}\) & \(\textcolor{blue}{1.232}\) & \(\textcolor{blue}{2.507}\) & \(\textcolor{blue}{0.593}\) & \(\textcolor{red}{-2.338}\) & \(\textcolor{blue}{1.241}\) & \(\textcolor{blue}{1.524}\) & \(\textcolor{blue}{0.427}\) & \(\textcolor{red}{-0.808}\) & \(\textcolor{black}{-0.009}\) & \(\textcolor{blue}{0.983}\) & \(\textcolor{black}{0.166}\)\\
\bottomrule
\end{tabular}}
\end{table}

\paragraph{Baseline Intensities} Table~\ref{tab:kappa-top5-lamp} lists the five largest baseline intensities per response group. The same five transitions appear in both groups with different order and magnitude. Transitions associated with obtaining the authorization number and submitting the return form dominate in both groups. The highest-intensity transitions thus overlap more between groups than in CD Tally. Response-group differences in routing are examined in Section~\ref{sec:lampcross}.

\begin{table}[htb!]
\centering
\caption{Five largest posterior mean baseline hazards $\kappa_{c,m,l}$ by response group for Lamp Return with 95\% HPD intervals. Log-scale summaries are computed from posterior draws of $\log\kappa$.}
\label{tab:kappa-top5-lamp}
\begin{threeparttable}
\small
\begin{tabularx}{\textwidth}{
@{}
l
>{\raggedright\arraybackslash}X
>{\raggedright\arraybackslash}X
>{\raggedleft\arraybackslash}p{3.5cm}
>{\raggedleft\arraybackslash}p{3.75cm}
@{}}
\toprule
Group & From & To & $\kappa$ mean [95\% HPD] & $\log(\kappa)$ mean [95\% HPD] \\
\midrule
\multirow{5}{*}{Correct $(c=1)$} & \texttt{wb\_pg\_8\_3} & \texttt{wb\_pg\_8\_3\_1} & 1867.7 [1633.1, 2083.4] & 7.53 [7.41, 7.65] \\
 & \texttt{wb\_pg\_8\_4\_submit} & \texttt{wb\_pg\_pop3} & 1630.3 [1397.7, 1869.1] & 7.39 [7.25, 7.54] \\
 & \texttt{wb\_pg\_8\_2} & \texttt{wb\_pg\_8} & 1149.7 [919.6, 1360.5] & 7.04 [6.84, 7.23] \\
 & \texttt{em\_move\_ok} & \texttt{em\_m\_move} & 553.4 [297.9, 825.8] & 6.28 [5.79, 6.78] \\
 & \texttt{paste} & \texttt{keypress} & 304.1 [237.3, 376.1] & 5.71 [5.47, 5.93] \\
\addlinespace
\multirow{5}{*}{Incorrect $(c=0)$} & \texttt{wb\_pg\_8\_4\_submit} & \texttt{wb\_pg\_pop3} & 1218.7 [1057.7, 1386.9] & 7.10 [6.96, 7.23] \\
 & \texttt{wb\_pg\_8\_2} & \texttt{wb\_pg\_8} & 1073.6 [856.2, 1306.2] & 6.97 [6.78, 7.20] \\
 & \texttt{wb\_pg\_8\_3} & \texttt{wb\_pg\_8\_3\_1} & 883.7 [786.7, 983.0] & 6.78 [6.67, 6.89] \\
 & \texttt{em\_move\_ok} & \texttt{em\_m\_move} & 793.0 [326.4, 1303.5] & 6.62 [5.98, 7.27] \\
 & \texttt{paste} & \texttt{keypress} & 467.9 [392.3, 550.2] & 6.14 [5.97, 6.31] \\
\bottomrule
\end{tabularx}
\begin{tablenotes}
\footnotesize
\item \textit{Note.}
Rows are ordered by posterior mean within each response group. $\kappa$ denotes the baseline transition hazard. Log-scale summaries were computed from posterior draws of $\log(\kappa)$, not by taking the logarithm of the posterior mean of $\kappa$.
\end{tablenotes}
\end{threeparttable}
\end{table}

\subsubsection{Transition-Network Comparisons Across Countries and Response Groups}\label{sec:lampcross}

\paragraph{Embedding and Distances} The same procedure was applied to the 28 Lamp Return networks with $d=32$, $d_1=64$, $\gamma^{+}=0.3$, and $\gamma^{-}=0.6$. Across the 280,000 posterior network draws the fixed encoder reduced the support-restricted KL divergence relative to the uniform baseline by 94.8\% on average (range 92.6\% to 96.0\% across combinations) with a mean NDCG@5 of 0.992. Full diagnostics are given in the Section~S3.3 of the Supplementary Material.

\begin{figure}[htb!]
    \centering
    \includegraphics[width=0.75\linewidth]{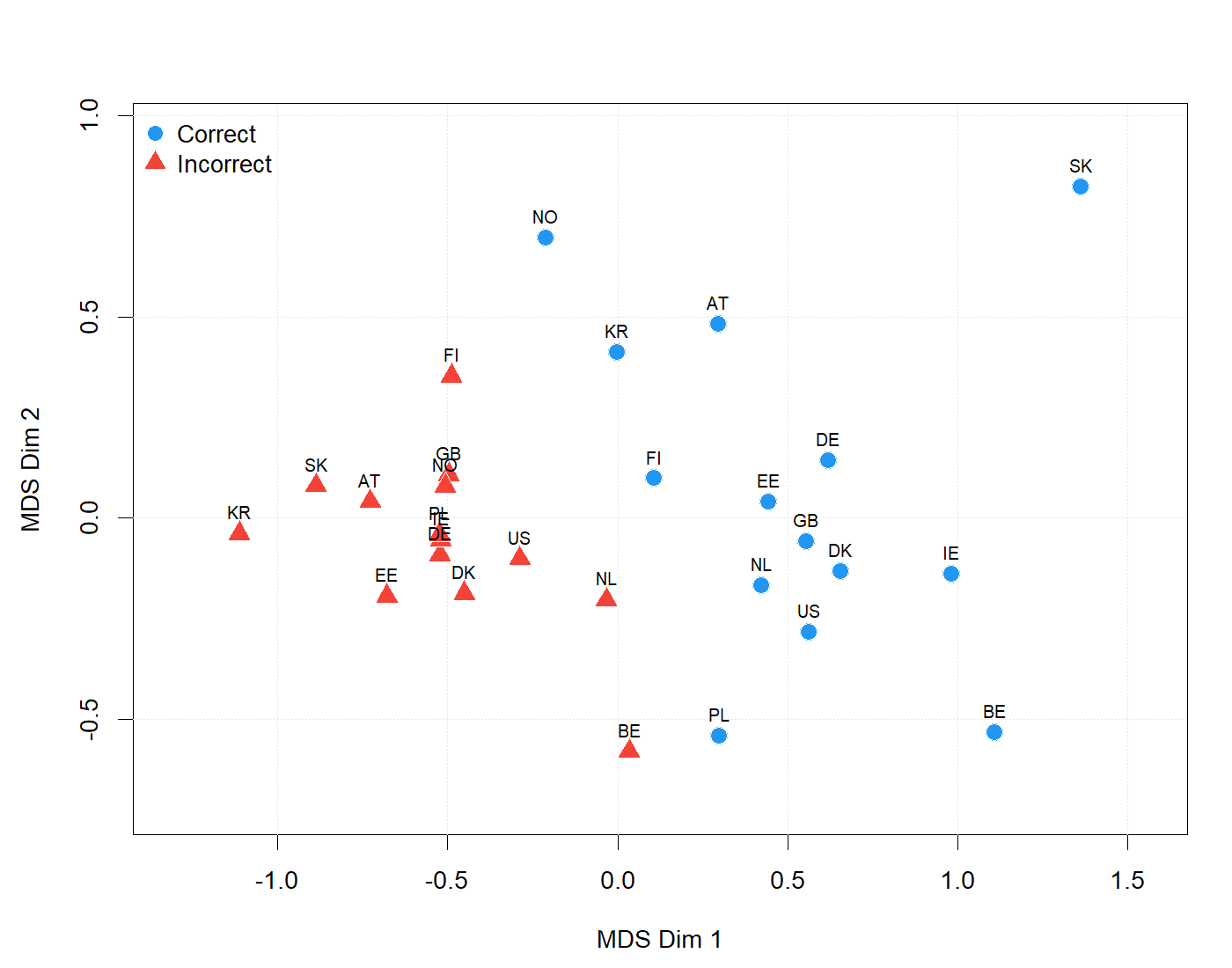}
    \caption{nMDS representation of the 28 Lamp Return transition networks based on the posterior mean Wasserstein distance matrix. Circles denote correct-response networks and triangles denote incorrect-response networks. Stress-1 = 0.134.}
    \label{fig:lamp_mds}
\end{figure}

\begin{figure}[htb!]
    \centering
    \begin{subfigure}[t]{0.43\linewidth}
        \centering
        \includegraphics[width=\linewidth]{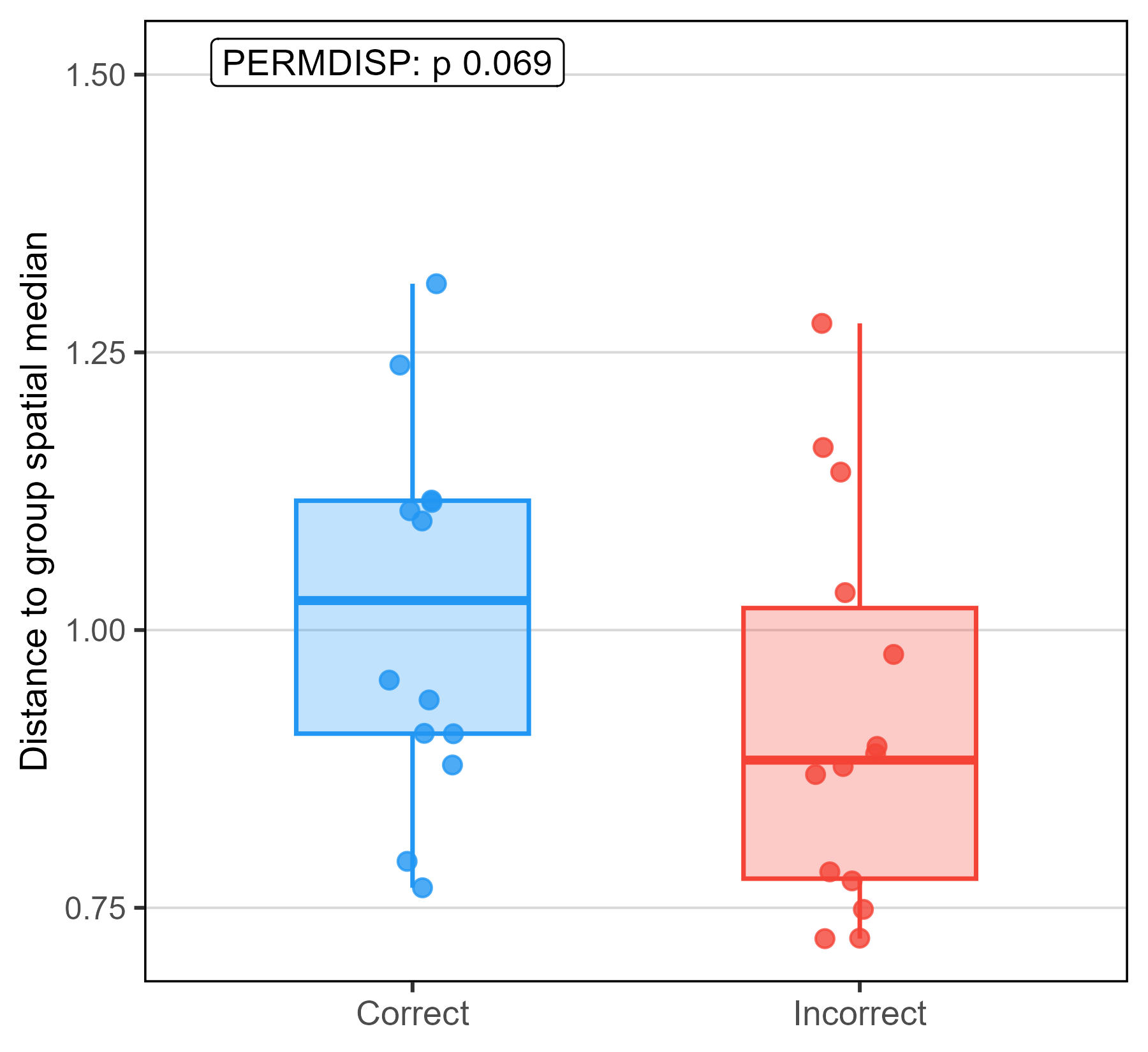}
        \caption{}
        \label{fig:lamp_dispersion_group}
    \end{subfigure}
    \hfill
    \begin{subfigure}[t]{0.4\linewidth}
        \centering
        \includegraphics[width=\linewidth]{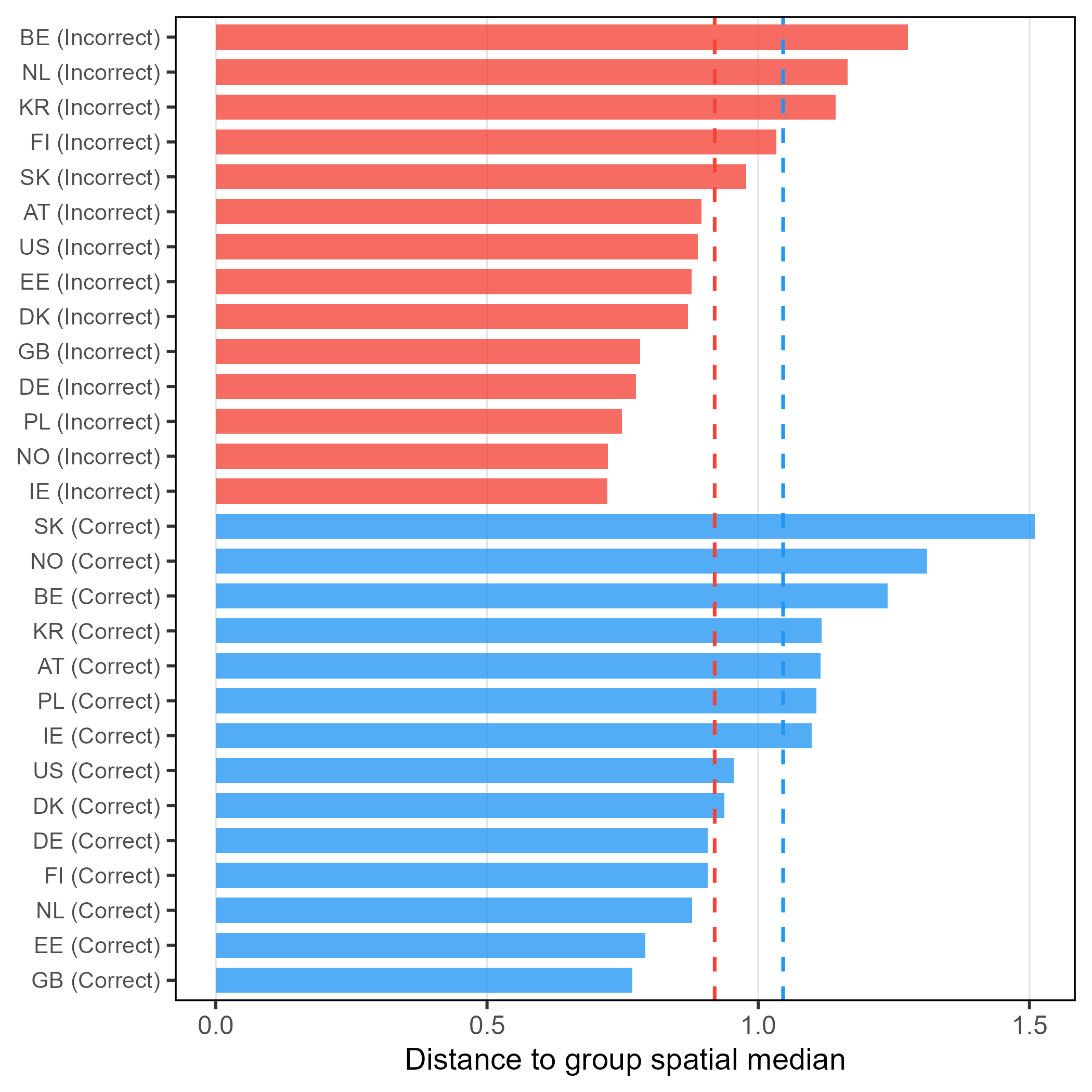}
        \caption{}
        \label{fig:lamp_dispersion_graph}
    \end{subfigure}
    \caption{Dispersion of the Lamp Return transition networks by response group. (a) Distances from country-specific networks to their response-group spatial median. (b) Network-specific distances to the corresponding group spatial median. Dashed lines indicate response-group means.}
    \label{fig:lamp_dispersion}
\end{figure}

\paragraph{Global Differences} Figure~\ref{fig:lamp_mds} shows the nMDS solution (Stress-1 = 0.134). Correct and incorrect networks occupy partly distinct regions. PERMANOVA indicated a difference between the response groups ($F(1,26)=7.353$, $R^2=0.220$, $p<0.001$). Unlike CD Tally PERMDISP did not indicate a difference in within-group dispersion ($F(1,26)=2.975$, $p=0.069$; Figure~\ref{fig:lamp_dispersion}). The two groups thus differ in location in the network-distance space whereas the difference in within-group dispersion was not statistically significant.

\paragraph{Local Structure} We applied the AME decomposition to the correct and incorrect networks of GB and to the correct networks of SK and GB where SK lies apart from the main cluster of the correct group in the nMDS solution. Receiver and multiplicative effects are reported in the Section~S5 of the Supplementary Material.
\begin{itemize}
\item {\bf United Kingdom: Correct versus Incorrect} Among key actions \texttt{wb\_pg\_8\_4} (opening the return form) has the largest positive receiver effect toward the correct network ($b_l=0.078$) followed by \texttt{wb\_pg\_8\_4\_reason\_4} (selecting the wrong-item return reason; $b_l=0.070$). The action \texttt{em\_m\_view\_305} (viewing the email with the authorization number) also has a positive effect ($b_l=0.044$), whereas \texttt{em\_f\_view} has a small negative effect ($b_l=-0.019$). Two branching points carry the multiplicative differences. From the return-reason state \texttt{wb\_pg\_8\_4\_reason\_6} the transition to \texttt{wb\_pg\_8\_4\_reason\_4} is correct-oriented ($u_m^\top v_l=0.553$, $\Delta_{ml}=0.632$) whereas the transition to \texttt{wb\_pg\_8\_4\_request\_1} is incorrect-oriented ($u_m^\top v_l=-0.361$, $\Delta_{ml}=-0.375$). \texttt{wb\_pg\_8\_4\_request\_1} is itself the task-relevant exchange request and its receiver effect is relatively small ($b_l=0.011$). The pathway difference highlights how respondents proceed from the preceding reason state. After viewing the authorization email the transition \texttt{em\_m\_view\_305} $\rightarrow$ \texttt{wb} is correct-oriented ($u_m^\top v_l=0.226$, $\Delta_{ml}=0.404$) whereas \texttt{em\_m\_view\_305} $\rightarrow$ \texttt{keypress} is incorrect-oriented ($u_m^\top v_l=-0.200$, $\Delta_{ml}=-0.190$). Support also differs. From the incorrect reason \texttt{wb\_pg\_8\_4\_reason\_1} the correct network contains a transition to \texttt{wb\_pg\_8\_4\_reason\_4} whereas the incorrect network contains a transition directly to \texttt{wb\_pg\_8\_4\_request\_1}. The correct network thus contains a correction of an inappropriate intermediate choice that the incorrect network lacks.

\item {\bf Slovakia versus United Kingdom within the Correct Group} The differences are small. Among key actions \texttt{wb\_pg\_8} has the largest positive receiver effect toward SK ($b_l=0.036$) and \texttt{wb\_pg\_8\_4\_reason\_4} a smaller one ($b_l=0.009$) whereas \texttt{em\_f\_view} ($b_l=-0.016$) and \texttt{wb\_pg\_8\_4\_request\_1} ($b_l=-0.013$) are oriented toward GB. The transition \texttt{wb\_pg\_8\_4\_reason\_6} $\rightarrow$ \texttt{wb\_pg\_8\_4\_reason\_4} is stronger in SK ($u_m^\top v_l=0.141$, $\Delta_{ml}=0.199$) and the direct pathway from \texttt{wb\_pg\_8\_4\_reason\_6} to \texttt{wb\_pg\_8\_4\_request\_1} is observed only in GB. Because both networks represent correct responses these patterns reflect different intermediate routes to a successful solution. The core key-action structure is largely shared. Of the 85 key-to-key pathways observed in the union of the two networks 70 (82.4\%) are present in both and the largest absolute difference among shared pathways is $|\Delta_{ml}|=0.022$ with the largest multiplicative component $|u_m^\top v_l|=0.030$. Cross-country differences among correct respondents thus lie in the routes into the task-relevant structure rather than in the connections among key actions.
\end{itemize}
\section{Discussion}\label{sec:discussion}

Process data make it possible to study not only whether respondents solve a task, but also how they move through it and how quickly those transitions occur. The present analysis shows that these different aspects of the response process need not tell the same story. Across the two PIAAC tasks, countries differed in transition pace, respondent characteristics showed systematic associations with transition intensity, and correct and incorrect respondents differed in the global organization of their transition networks. At the same time, the form and extent of these differences depended on the task. These findings illustrate the value of treating timing and routing as distinct but related components of the response process rather than reducing process data to a single summary such as total response time or final accuracy.

The statistical framework was developed to address several features of these data that arise together in the PIAAC application. Country-by-response-group samples are unequal, many transitions are rare, and the sets of observed transitions differ across groups. The hierarchical multi-state survival model addresses the first two features by partially pooling country-specific effects while estimating between-country heterogeneity directly. The network component addresses the third by representing posterior transition probabilities as directed weighted networks and comparing their structure in a common latent space despite heterogeneous supports. Evaluating Wasserstein distances across posterior draws further carries uncertainty from the fitted transition model into the global network analysis. The combination therefore separates three features of the data that would otherwise be difficult to study jointly: transition pace and intensity, between-country heterogeneity, and the routing of respondents among possible next actions.

The empirical results demonstrate why this separation is useful. Age and Eskill showed the most consistent associations with transition intensity across countries and across both items: older respondents tended to move more slowly through the tasks, whereas respondents with greater computer and Internet experience tended to move more quickly. Country-level mean speed also varied appreciably. These findings concern the \emph{pace} of task performance, however, and do not by themselves describe how respondents navigate the task. The transition-network analysis revealed additional structure that was not captured by the speed and covariate effects.

For CD Tally, the correct-response networks were considerably more concentrated across countries than the incorrect-response networks. One interpretation is that successful performance on this task tends to involve a relatively common routing structure centered on the task-relevant spreadsheet operations, whereas unsuccessful performance can arise through a wider range of paths. This pattern was not equally apparent for Lamp Return. Correct and incorrect networks differed in their global location, but the evidence for a difference in cross-country dispersion was weaker. The difference between the two items is substantively informative. Lamp Return involves a larger action space and requires respondents to coordinate information across email and web environments, and its incorrect-response group combines scores from 0 to 2. Consequently, respondents classified as incorrect may include both those who made little progress and those who completed several task-relevant steps without receiving full credit. The observed heterogeneity in process structure therefore depends not only on response correctness but also on the architecture and scoring of the task.

The key-action effects provide a second example of this task dependence. For CD Tally, the correct and incorrect groups differed primarily in the source- and destination-key effects: correct respondents were less likely to move away from key actions toward non-key actions and more likely to move into key actions. For Lamp Return, both response groups showed the same general pattern of key-action effects, and the distinction between groups was more strongly expressed through the source effect and the interaction between source and destination key-action status. Thus, treating an action simply as ``important'' or ``unimportant'' would obscure meaningful differences in how it functions within a response sequence. Distinguishing whether a key action serves as the origin of a transition, the destination of a transition, or part of a key-to-key transition provides a more informative description of the response process.

The local network analyses further show that global network differences can often be traced to a small number of interpretable branching points. In CD Tally, the Austrian correct and incorrect networks differed in what respondents did after confirming the spreadsheet sorting options. In Lamp Return, the United Kingdom networks differed in whether respondents revised the reason for the return and in what they did after viewing the authorization email. By contrast, the correct networks for Slovakia and the United Kingdom shared most of their key-action connections and differed mainly in the surrounding routes into those actions. These examples illustrate an important feature of process data: the interpretation of an action depends on the path through which it is reached and on what follows it. Global network summaries identify where response processes differ, whereas local decompositions help identify the specific transitions responsible for those differences.

In summary, the results suggest that cross-national process data are most informative when examined at multiple resolutions. Country differences in overall transition pace need not coincide with differences in routing, and similar final outcomes need not imply similar response processes. Conversely, respondents from different countries can reach the same successful outcome through somewhat different intermediate routes while sharing a common core of task-relevant actions. The framework therefore complements conventional cross-national comparisons of proficiency or accuracy by providing information about the processes through which those outcomes are produced. Such information may be useful for understanding task functioning, identifying common and atypical solution pathways, and generating hypotheses about why a task operates differently across groups. Because the analysis is observational, however, country differences should be interpreted descriptively rather than as effects of national or cultural context.

Several modeling choices delimit these conclusions. Transition-pair baseline intensities are shared across countries within each response group. This stabilizes estimation for sparse transitions but places cross-country variation primarily in the hierarchical covariate, key-action, and speed effects; allowing country-specific deviations in the baseline intensities would provide a more flexible alternative when larger samples are available. The time-homogeneous Markov specification conditions on the current action only, so dependence on longer response histories is not represented. History-dependent transition effects could be incorporated in future extensions.

The network analysis also conditions on the observed transition support. Unobserved transitions are assigned zero probability, although absence from a finite sample does not imply that a transition is impossible. Some apparent support differences may therefore reflect limited opportunities to observe rare transitions rather than genuine differences in response processes. Similarly, the key-action sets are selected from their association with response correctness in the same data, so substantive interpretations are conditional on that selection procedure. External or task-design-based definitions of key actions would provide a useful sensitivity analysis.

Finally, posterior uncertainty is propagated from the multi-state model into the pairwise Wasserstein distances, but the subsequent nMDS, PERMANOVA, PERMDISP, and AME analyses are based on posterior summaries. Extending uncertainty propagation through these downstream analyses would provide a more fully Bayesian treatment. The network results also depend on the learned representation. In the present analyses, reconstruction diagnostics indicated that dominant transitions were well preserved and the Wasserstein distances were strongly consistent in rank with alternative distributional dissimilarities, but alternative encoders may induce somewhat different network geometries. Developing joint hierarchical models for local network structure across all countries, rather than examining selected pairwise AME contrasts, is another natural direction for future work.

More broadly, the statistical issues motivating this analysis are not unique to educational assessment. Similar data structures arise whenever individuals move through a sequence of discrete states over continuous time and the transition systems are sparse and heterogeneous across groups. The present framework provides one way to combine hierarchical event-history modeling with global and local network representations in such settings. In the PIAAC application, this combination reveals variation at three complementary levels: how quickly respondents move through a task, how their overall routing structures differ, and which local transitions account for those differences. These distinctions would be difficult to recover from final outcomes or from either the survival or network representation alone.

\section*{Acknowledgments}

This work was partially supported by the National Research Foundation of Korea [grant number RS-2023-00217705, and RS-2024-00333701; Basic Science Research Program awarded to I.H.J.], the ICAN (ICT Challenge and Advanced Network of HRD) support program [grant number RS-2023-00259934, awarded to I.H.J.], supervised by the IITP (Institute of Information \& Communications Technology Planning \& Evaluation), and the Ministry of Trade, Industry, and Energy (MOTIE), Korea, under the project ``Industrial Technology Infrastructure Program'' [RS-2024-00466693, awarded to I.H.J.]. Correspondence should be addressed to Ick Hoon Jin, Department of Applied Statistics, Department of Statistics and Data Science, Yonsei University, Seoul, Republic of Korea. E-Mail: ijin@yonsei.ac.kr. 

\section*{Data Availability}
The PIAAC log data analyzed in this study are available through GESIS at \url{https://doi.org/10.4232/1.12955}. Preprocessing followed the procedures described by \citet{park2025analysis}, with additional processing to combine the country-specific datasets for the joint analysis. Code for this additional processing, model estimation, and network analysis is available at \url{https://github.com/Leeju1/HMSM}.
% The datasets generated during and/or analyzed during the current study are available in the GESIS, \url{https://doi.org/10.4232/1.12955}. The preprocessing scripts, sampler code, and simulation files accompanying this paper are maintained at \url{https://github.com/P-JuNYeonG/action-irt}.

\section*{AI Use Statement}
GPT-6 Astra (medium reasoning setting) and Claude Fable 5.1 were used to check for typographical and grammatical errors. The authors reviewed all suggested corrections and take full responsibility for the content of the manuscript.
%Large language models were used at three points in the preparation of this manuscript. Claude Sonnet 4.5 was used for the description standardization step of the preprocessing pipeline, as described in Sections~\ref{sec:preprocessing_pipeline} and~\ref{sec:llm_preprocessing}. Figure~\ref{fig:framework} was drafted with ChatGPT (GPT-5.5, medium reasoning setting). Claude Opus 5.0 was used to check for typographical and grammatical errors. The authors reviewed all output and take full responsibility for the content of the manuscript.

\bibliographystyle{apalike}
\bibliography{reference}
\end{document}